\documentclass[12pt]{article}
\usepackage{latexsym,amsmath,amssymb,mathrsfs, graphicx}
\renewcommand{\thefootnote}{\fnsymbol{footnote}}

\usepackage{amsmath}
\usepackage{amsfonts}

\numberwithin{equation}{section}

\def\doubleset#1#2{\bgroup%
\def\doit#1#2{%
\setbox\dblsetbox=\hbox{$\cstyle #1$}%
\raise#2\ht\dblsetbox\copy\dblsetbox%
\hskip-\wd\dblsetbox%
\raise-#2\ht\dblsetbox\box\dblsetbox}%
\mathchoice%
{\def\cstyle{\displaystyle}\doit#1#2}%
{\def\cstyle{\textstyle}\doit#1#2}%
{\def\cstyle{\scriptstyle}\doit#1#2}%
{\def\cstyle{\scriptscriptstyle}\doit#1#2}\egroup}
\def\underarrow#1{\vbox{\ialign{##\crcr$\hfil\displaystyle
 {#1}\hfil$\crcr\noalign{\kern1pt\nointerlineskip}$\longrightarrow$\crcr}}}

\newbox\dblsetbox

\newlength{\extraspace}
\newlength{\extraspaces}
\newcommand{\be}{\begin{equation}
\addtolength{\abovedisplayskip}{\extraspaces}
\addtolength{\belowdisplayskip}{\extraspaces}
\addtolength{\abovedisplayshortskip}{\extraspace}
\addtolength{\belowdisplayshortskip}{\extraspace}}
\newcommand{\ee}{\end{equation}}

\newcommand{\ba}{\begin{eqnarray}
\addtolength{\abovedisplayskip}{\extraspaces}
\addtolength{\belowdisplayskip}{\extraspaces}
\addtolength{\abovedisplayshortskip}{\extraspace}
\addtolength{\belowdisplayshortskip}{\extraspace}}
\newcommand{\ea}{\end{eqnarray}}

\newcommand{\bd}{\begin{displaymath}
\addtolength{\abovedisplayskip}{\extraspaces}
\addtolength{\belowdisplayskip}{\extraspaces}
\addtolength{\abovedisplayshortskip}{\extraspace}
\addtolength{\belowdisplayshortskip}{\extraspace}}
\newcommand{\ed}{\end{displaymath}}

\newcounter{saveeqn}

\newcommand{\newsection}[1]{
\vspace{12mm}
\pagebreak[3]
\addtocounter{section}{1}
\setcounter{equation}{0}
\setcounter{subsection}{0}
\noindent{\bf \thesection. #1}
\nopagebreak
\medskip
\nopagebreak}

\newcommand{\newsubsection}[1]{
\vspace{0.8cm}
\pagebreak[3]
\addtocounter{subsection}{1}
\noindent{\it \thesubsection. #1}
\nopagebreak
\vspace{2mm}
\nopagebreak}

\begin{document}
\addtolength{\baselineskip}{1.5mm}

\thispagestyle{empty}
\vbox{}
\vspace{2.5cm}

\begin{center}
\centerline{\LARGE{Equivariant Cohomology Of The Chiral de Rham Complex}}
\bigskip
\centerline {\LARGE{And The Half-Twisted Gauged Sigma Model}}

\vspace{2.5cm}

{Meng-Chwan~Tan\footnote{E-mail: phytmc@nus.edu.sg}}
\\[0mm]
{\it Department of Physics\\
National University of Singapore \\
Singapore 119260}\\[8mm]
\end{center}

\vspace{0.5cm}

\centerline{\bf Abstract}\bigskip \noindent

In this paper, we study the perturbative aspects of the half-twisted variant of Witten's topological A-model coupled to a non-dynamical gauge field with K\"ahler target space $X$ being a $G$-manifold.    Our main objective is to furnish a purely physical interpretation of the equivariant cohomology of the chiral de Rham complex,   recently constructed by Lian and Linshaw in \cite{andy1},  called the ``chiral equivariant cohomology''. In doing so, one finds that key mathematical results such as the vanishing in the chiral equivariant cohomology of positive weight classes, lend themselves to straightforward physical explanations. In addition, one can also construct topological invariants of $X$ from the correlation functions of the  relevant physical operators corresponding to the non-vanishing weight-zero classes. Via the topological invariance of these correlation functions, one can verify, from a purely physical perspective,  the mathematical isomorphism between the weight-zero subspace of the chiral equivariant cohomology and the classical equivariant cohomology of $X$. Last but not least, one can also determine fully, the de Rham cohomology ring of $X/G$, from the topological chiral ring generated by the local   ground operators of the physical model under study.

\newpage
\renewcommand{\thefootnote}{\arabic{footnote}}
\setcounter{footnote}{0}

\newsection{Introduction}

The mathematical theory of the Chiral de Rham complex or CDR for short, was first introduced in two seminal papers \cite{MSV1, MSV2} by Malikov et al. in 1998. It aims to provide a rigorous mathematical construction of conformal field theories in two-dimensions without  resorting to mathematically non-rigorous methods such as the path integral. Since its introduction, the CDR has found many interesting applications in various fields of geometry and representation theory, namely mirror symmetry \cite{Bo}, and the study of elliptic genera \cite{BL, BL1, BL2}. It is by now a fairly well-studied object in the mathematical literature.

Efforts to provide an explicit physical interpretation of the theory of CDR were undertaken in \cite{Ka, MC}. In essence, one learns that the local sections of the sheaf of CDR on a manifold with complex dimension $n$, can be described by a holomorphic $N=2$ SCFT which is a tensor product of $n$ copies of the holomorphic $bc$-$\beta\gamma$ system: the space of sections is simply the algebra of local operators graded by their ghost numbers and conformal weights. Alternatively, one can also deduce this interpretation from the mathematical definition of the sheaf of CDR on an affine space \cite{MSV1, MSV2}.

The CDR is also an example of what is mathematically known as a differential vertex algebra. By synthesizing the algebraic approach to classical equivariant cohomology with the theory of differential vertex algebras, and using an appropriate notion of invariant theory (also known as the coset construction in physics), Lian and Linshaw   recently  constructed,  on any $G$-manifold $X$, an equivariant cohomology of the CDR called the chiral equivariant cohomology \cite{andy1}. This new equivariant cohomology theory was also developed further in a second paper \cite{andy2}, where several interesting mathematical results such as the vanishing of positive weight classes (when $X$ is not a point) were established.

In this paper, we explore the half-twisted A-model coupled to a non-dynamical gauge field with gauge group $G$ and K\"ahler target space $X$.  The main objective is to furnish a purely  physical interpretation of  the chiral equivariant cohomology. In doing so, we hope to obtain straightforward physical explanations of some of the established mathematical results, and   perhaps,  even gain some novel insights into the physics via a reinterpretation of the known mathematics.

\smallskip\noindent{\it A Brief Summary and Plan of the Paper}

A brief summary and plan of the paper is as follows. In Section 2, we will start by first reviewing the construction and relevant features of the perturbative half-twisted A-model on any smooth $G$-manifold $X$, where $G$ is a compact group of automorphisms of $X$ which leave fix its metric and almost complex structure.

In Section 3, we will proceed to couple the model to a non-dynamical gauge field which takes values in the Lie algebra spanned by the vector fields generating the associated free  $G$-action   on $X$. Thereafter, we will discuss the pertinent features of the model which will be most relevant to our paper.

In Section 4, we specialise to the case when the gauge group $G$ is an abelian one such as $U(1)^d$ for any $d$. We then study what happens in the infinite-volume or weak-coupling limit. It is at this juncture that we first make contact with the chiral equivariant cohomology of \cite{andy1}. We then proceed to provide a straightforward physical explanation of a mathematical result in \cite{andy2} stating the vanishing in the chiral equivariant cohomology of positive weight classes. Next, we show that one can define a set of topological invariants on $X$ from the correlation functions of the relevant physical operators corresponding to non-trivial classes of the chiral equivariant cohomology. These correlation functions can in turn be used to furnish a    purely physical verification of the isomorphism  between the weight-zero subspace of the chiral equivariant cohomology and the classical equivariant cohomology of $X$ (as established in the mathematical literature in \cite{andy1,andy2}). Moreover, one can also determine fully, the de Rham cohomology ring of $X/G$,  from a topological chiral ring generated by the local   ground operators of the half-twisted gauged sigma model.   Last but not least, we show that our results hold in the large but finite-volume limit as well, that is, to all orders of perturbation theory.

In Section 5, we conclude the paper with a discussion of some open problems that we hope to address in a future publication.

\newsection{The Half-Twisted  A-Model on a Smooth $G$-Manifold $X$}

In this section, we will review the construction and relevant features of the perturbative half-twisted A-model on a smooth K\"ahler manifold $X$. For the purpose of our paper, we will implicitly assume that $X$ is a smooth $G$-manifold. In other words, one can define a free $G$-action on $X$, which in our case, will be generated by a set of vector fields (on $X$) which furnish a Lie algebra $\mathfrak g$  of $G$. The review in this section is to serve as a prelude to section 3, where we will discuss  the construction of the half-twisted gauged A-model on $X$, our primary interest in this paper.

\newsubsection{The Construction of the Half-Twisted A-Model}

To begin with, let us first recall the half-twisted variant of the A-model in perturbation theory. It governs maps $\Phi : \Sigma \to X$, with $\Sigma$ being the worldsheet Riemann surface. By picking local coordinates $z$, $\bar z$ on $\Sigma$, and $\phi^{i}$, $\phi^{\bar i}$ on the K\"ahler manifold $X$, the map $\Phi$ can then be described locally via the functions $\phi^{i}(z, \bar z)$ and $\phi^{\bar i}(z, \bar z)$. Let $K$ and ${\overline K}$ be the canonical and anti-canonical bundles of $\Sigma$ (the bundles of one-forms of types $(1,0)$ and $(0,1)$ respectively), whereby the spinor bundles of $\Sigma$ with opposite chiralities are given by $K^{1/2}$ and ${\overline K}^{1/2}$. Let $TX$ and $\overline {TX}$ be the holomorphic and anti-holomorphic tangent bundle of $X$. The half-twisted variant as defined in \cite{n=2}, has the same classical Lagrangian as that of the original A-model in \cite{mirror manifolds}.\footnote{The action just differs from the A-model action in \cite{mirror manifolds} by a term $\int_{\Sigma} \Phi^*(K)$, where $K$ is the K\"ahler $(1,1)$-form on $X$. This term is irrelevant in perturbation theory where one considers only trivial maps $\Phi$ of degree zero.} (The only difference is that the cohomology of operators and states is taken with respect to a $\it{single}$ right-moving supercharge only instead of a linear combination of a left- and right-moving supercharge. This will be clear shortly). The action is thus given by
\begin{eqnarray}
S & = & \int_{\Sigma} |d^2z| \left( g_{i{\bar j}} \partial_z \phi^{\bar j} \partial_{\bar z}\phi^i + g_{i{\bar j}} \psi_{\bar z}^i D_z \psi^{\bar j} + g_{i \bar j} \psi^{\bar j}_z D_{\bar z} \psi^i  - R_{i {\bar k}  j {\bar l}} {\psi}^{i}_{\bar z} {\psi}^{\bar k}_z  \psi^j \psi^{\bar l}  \right),
\label{S}
\end{eqnarray}
where $|d^2z| = i dz \wedge d\bar z$ and $i,j,k, l = 1, 2, \dots, {\textrm{dim}_{\mathbb C} X}$. $R_{i {\bar k}  j {\bar l}}$ is the curvature tensor with respect to the Levi-Civita connection $\Gamma^i{}_{lj} = g^{i \bar k}\partial_lg_{j \bar k}$, and the covariant derivatives with respect to the connection induced on the worldsheet are given by
\be
D_z\psi^{\bar j} = \partial_z \psi^{\bar j} + \Gamma^{\bar j}{}_{\bar i \bar k} \partial_z\phi^{\bar i}\psi^{\bar k}, \qquad  D_{\bar z}\psi^{i} = \partial_{\bar z} \psi^{i} + \Gamma^{i}{}_{j k} \partial_{\bar z} \phi^{j}\psi^{k}.
\ee
The various fermi fields transform as smooth sections of the following bundles:
\begin{eqnarray}
\psi^i  \in  \Gamma \left(\Phi^*{TX}  \right), & \qquad & \psi^{\bar i}_{z}  \in  \Gamma \left( K \otimes \Phi^*{\overline{TX}}\right), \nonumber \\
\psi^i_{\bar z} \in  \Gamma \left({\overline K} \otimes \Phi^*{TX} \right), & \qquad &  \psi^{\bar i} \in  \Gamma \left(\Phi^*{\overline{TX}}\right), \\
\nonumber
\end{eqnarray}
Notice that we have included additional indices in the above fermi fields so as to reflect their geometrical characteristics on $\Sigma$; fields without a $z$ or $\bar z$ index transform as worldsheet scalars, while fields with a $z$ or $\bar z$ index transform as $(1,0)$ or $(0,1)$ forms on the worldsheet. In addition, as reflected by the $i$, and $\bar i$ indices, all fields continue to be valued in the pull-back of the corresponding bundles on $X$.

Note that the action $S$ in (\ref{S}) can be written as
\be
S = \int_{\Sigma}|d^2 z| \{Q, V \},
\label{Stop}
\ee
where
\be
V = i g_{i \bar j} ( \psi^i_{\bar z} \partial_z \phi^{\bar j} + \psi^{\bar j}_z \partial_{\bar z} \phi^i - {1\over 2} \psi^{\bar j}_z H^i_{\bar z} - {1\over 2} \psi^i_{\bar z}H^{\bar j}_z),
\label{V}
\ee
and $\delta V = - i \epsilon \{Q, V\}$, whereby  $\delta V$ is the variation of $V$ under the field transformations generated by the nilpotent BRST supercharge $Q$, which is given by $Q= Q_L + Q_R$. Here, $Q_L$ and $Q_R$ are left- and right-moving BRST supercharges respectively, and the field transformations generated by the supercharge $Q$ are given by
\begin{eqnarray}
\label{varfirst}
\delta\psi^j & = & 0, \\
\delta\psi^{\bar j} & = & 0,\\
\delta \phi^i & = & {\epsilon_+} \psi^i,\\
\delta \phi^{\bar i} & = & {\bar \epsilon}_- \psi^{\bar i}, \\
\delta \psi^i_{\bar z} & = & - {\bar \epsilon}_- H^i_{\bar z} - {\epsilon_+} \Gamma^i_{j k} \psi^j \psi^k_{\bar z},\\
\delta \psi^{\bar i}_{z}& = & - {{\epsilon_+}}H^{\bar i}_{z} - {\bar \epsilon}_- \Gamma^{\bar i}_{\bar j \bar k} \psi^{\bar j} \psi^{\bar k}_{z},\\
\label{h1}
\delta H^i_{\bar z}& = & R^i{}_{k \bar j l}\psi^k \psi^{\bar j} \psi^l_{\bar z} -  \Gamma^i_{jk} \psi^j H^k_{\bar z}, \\
\delta H^{\bar i}_{z} & = & R^{\bar i}{}_{\bar j l \bar k}\psi^{\bar j}\psi^l \psi^{\bar k}_{z} - \Gamma^{\bar i}_{\bar j \bar k} \psi^{\bar j} H^{\bar k}_{z}. \
\label{h2}
\label{varlast}
\end{eqnarray}
In the above, $\epsilon_+$ and ${\bar \epsilon}_-$ are $c$-number parameters associated with the BRST supersymmetries generated by $Q_L$ and $Q_R$. For notational simplicity, we have set $\epsilon_+$ and ${\bar \epsilon}_-$ in (\ref{h1}) and (\ref{h2}) to be 1. Note that we have used the equations of motion $H^i_{\bar z} = \partial_{\bar z}\phi^i$ and $H^{\bar i}_{z} = \partial_z \phi^{\bar i}$ to eliminate the auxillary fields $H^i_{\bar z}$ and $H^{\bar i}_{z}$ in our computation of (\ref{Stop}), so that we can obtain $S$ in (\ref{S}).

\newsubsection{Spectrum of Operators in the Half-Twisted A-Model}

As mentioned earlier, the half-twisted A-model is a greatly enriched variant in which one ignores $Q_L$ and considers $ Q_R$ as the BRST operator \cite{n=2}. Since the corresponding cohomology is now defined with respect to a single, right-moving, scalar  supercharge $Q_R$, its classes need not be restricted to dimension $(0,0)$ operators (which correspond to ground states). In fact, the physical operators will have dimension $(n,0)$, where $n \geq 0$. Let us verify this important statement.

From (\ref{S}), we find that the anti-holomorphic stress tensor takes the form $ T_{\bar z \bar z} =g_{i \bar j}  \partial_{\bar z} \phi^i \partial_{\bar z} \phi^{\bar j} + g_{i \bar j}  \psi_{\bar z}^i \left ( \partial_{\bar z} \psi^{\bar j} + \Gamma^{\bar j}_{\bar l \bar k}\partial_{\bar z} \phi^{\bar l} \psi^{\bar k} \right)$. One can go on to show that $ T_{\bar z \bar z} = \{ Q_R , i g_{i \bar j} \psi_{\bar z}^i \partial_{\bar z} \phi^{\bar j} \}$, that is, $T_{\bar z \bar z}$ is trivial in $Q_R$-cohomology. Now, we say that a local operator $\cal O$ inserted at the origin has dimension $(n,m)$ if under a rescaling $z\to \lambda z$, $\bar z\to \bar\lambda z$, it transforms as $\partial^{n+m}/\partial z^n\partial\bar z^m$, that is, as $\lambda^{-n}\bar\lambda{}^{-m}$. Classical local
operators have dimensions $(n,m)$ where $n$ and $m$ are non-negative integers.\footnote{Anomalous dimensions under RG flow may shift the values of $n$ and $m$ quantum mechanically, but the spin given by $(n-m)$, being an intrinsic property, remains unchanged.} However, only local operators with $m = 0$ survive in $Q_R$-cohomology. The reason for the last statement is that the rescaling of $\bar z$ is generated by $\bar L_0=\oint d\bar z\, \bar z T_{\bar z\bar z}$.  As we noted above, $T_{\bar z\,\bar z}$ is of the form $\{{Q}_R,\dots\}$, so $\bar L_0=\{{Q}_R,V_0\} $ for some $V_0$. If $\cal O$ is to be admissible as a local physical operator, it must at least be true that $\{{Q}_R, {\cal O}\}=0$. Consequently, $[\bar L_0,{\cal O}]=\{{Q}_R,[V_0,{\cal O}]\}$.  Since the eigenvalue of $\bar L_0$ on $\cal O$ is $m$, we have $[\bar L_0,{\cal O}]=m{\cal O}$. Therefore, if $m\not= 0$, it follows that ${\cal O}$ is $Q_R$-exact and thus trivial in $Q_R$-cohomology. On the other hand, the holomorphic stress tensor is given by $T_{zz} =   g_{i \bar j} \partial_z \phi^i \partial_z \phi^{\bar j} + g_{i \bar j} \psi^{\bar j}_z D_z \psi^i$, and one can verify that it can be written as  $T_{zz} = \{ Q_L , i g_{i \bar j} \psi^{\bar j}_z \partial_z \phi^i \}$,  that is, it is $Q_L$-exact. Since we are only interested in $Q_R$-closed modulo $Q_R$-exact operators, there is no restriction on the value that $n$ can take. These arguments persist in the quantum theory, since a vanishing cohomology in the classical theory continues to vanish when quantum effects are small enough in the perturbative limit.

Hence, in contrast to the A-model, the BRST spectrum of physical operators and states in the half-twisted model is infinite-dimensional. A specialisation of its genus one partition function, also known as the elliptic genus of $X$, is given by the index of the $Q_R$ operator. Indeed, the half-twisted  model is not a topological field theory, rather, it is a 2d conformal field theory - the full stress tensor derived from its action is exact with respect to the  combination $Q_L +Q_R$, but not $Q_R$ alone.

\newsubsection{The Ghost Number Anomaly}

Let us now touch upon a particular symmetry of the action $S$ which will be relevant to our study. Note that $S$ has a left and right-moving ``ghost number" symmetry whereby the left-moving fermionic fields transform as $\psi^i \to e^{i\alpha}\psi^i$ and $\psi^{\bar i}_{z} \to e^{-i \alpha} \psi^{\bar i}_{z}$, while the right-moving fermionic fields transform as $\psi^{\bar i} \to e^{i \alpha}\psi^{\bar i}$ and $\psi^i_{\bar z} \to e^{-i \alpha}\psi^i_{\bar z}$, where $\alpha$ is real. In other words, the fields $\psi^i$, $\psi^{\bar i}_{ z}$, $\psi^{\bar i}$ and $\psi^i_{\bar z}$ can be assigned the $(g_L, g_R)$ left-right ghost numbers $(1,0)$, $(-1,0)$, $(0,1)$ and $(0,-1)$ respectively. However, there is a ghost number anomaly at the quantum level, and one will need to place some restrictions on the form that the physical operators in the $Q_R$-cohomology can take, if there is to be a cancellation of this anomaly. As an example, let us consider a general, dimension $(0,0)$ operator  ${\cal O}^{p_i, q_i}(z, \bar z) = A(\phi^j, \phi^{\bar j})_{k_1, k_2, \dots k_{p_i}, {\bar l}_1, {\bar l}_2, \dots, {\bar l}_{q_i}} \psi^{k_1} \psi^{k_2} \dots \psi^{k_{p_i}} \psi^{\bar l_1} \psi^{\bar l_2} \dots \psi^{\bar l_{q_i}}$ of ghost number $(p_i, q_i)$ which is in the $Q_R$-cohomology. Let the correlation function of $\it{s}$ such operators be $Z = <{\cal O}^{p_1, q_1}{\cal O}^{p_2, q_2} \dots {\cal O}^{p_s, q_s} >$. Via the Hirzebruch-Riemann-Roch theorem, we find that one must have
\be
\sum^{\it{s}} _{i=1} p_i = \sum^s_{i=1} = q_i = \int_{\Sigma}\Phi^*c_1(TX) + \textrm{dim}_{\mathbb C} X(1-g)
\label{index}
\ee
or $Z$ will vanish. Here, $g$ is the genus of the worldsheet Riemann surface $\Sigma$. In perturbation theory, one considers only  degree-zero maps $\Phi$. Thus, the first term on the RHS of (\ref{index}) will vanish in our case. Since $p_i$ and $q_i$ correspond respectively to the number of $\psi^j$ and $\psi^{\bar j}$ fields in the operator ${\cal O}^{p_i, q_i}$, they cannot take negative values. Hence, in order to have a consistent theory, we see from (\ref{index}) that $\Sigma$ must be of genus-zero. In other words, the relevant worldsheet is a simply-connected Riemann surface in perturbation theory.

\newsubsection{Reduction from $N=1$ Supersymmetry in 4d}

Note that in order to untwist the A-model, one needs to restore the $SO(2)$ rotation generator of the 2d theory. This amounts to a redefinition of the worldsheet  spins of the fermionic fields $\psi^j$, $\psi^{\bar j}$, $\psi^k_{\bar z}$ and $\psi^{\bar k}_z$ so that they will transform as worldsheet spinors again.\footnote{To twist an $N=(2,2)$ supersymmetric sigma model into an A-model, we start with the Euclidean version of the theory from the Minkowski theory by a Wick rotation of the coordinates first. This means that the $SO(1,1)$ Lorentz group is now the Euclidean rotation group $SO(2)_E$. We then `twist' the theory by replacing the rotation generator $M_E$ of the $SO(2)_E$ group with $M'_E = M_E + F_V$, where $F_V$ is the generator of the vector R-symmetry of the theory. This is equivalent to redefining the spins of the various fields as $s' = s + {q_V \over 2}$, where $s$ is the original spin of the field, and $q_V$ is its corresponding vector R-charge.}  In short, one must make the replacements $\psi^j \to \psi^j_-$, $\psi^{\bar j} \to \psi^{\bar j}_+$, $\psi^j_{\bar z} \to \psi^j_+$ and $\psi^{\bar j}_z \to \psi^{\bar j}_-$, where the $-$ or $+$ subscript indicates that the corresponding field transforms  as a section of the bundle $K^{1/2}$ or ${\overline K}^{1/2}$ respectively on $\Sigma$. In addition, as before, a $j$ or $\bar j$ superscript also indicates that the relevant field in question will take    values in the pull-back of $TX$ or $\overline{TX}$. The form of the resulting, untwisted action is similar to (\ref{S}), and it is just the action of an $N=(2,2)$ supersymmetric non-linear sigma model in two-dimensions:
\begin{eqnarray}
S' & = & \int_{\Sigma} |d^2z| \left( g_{i{\bar j}} \partial_z \phi^{\bar j} \partial_{\bar z}\phi^i + g_{i{\bar j}} \psi_+^i D_z \psi^{\bar j}_+  + g_{i \bar j} \psi^{\bar j}_- D_{\bar z} \psi^i_-  - R_{i {\bar k}  j {\bar l}} {\psi}^{i}_+ {\psi}^{\bar k}_-  \psi^j_- \psi^{\bar l}_+  \right).
\label{S'}
\end{eqnarray}
The supersymmetric variation of the fields under which $S'$ is invariant read
\begin{eqnarray}
\label{first}
\delta \phi^i & = & \epsilon_+ \psi^j_+ +  \epsilon _- \psi^j_+,\\
\delta \phi^{\bar i} & = & {\bar \epsilon}_- \psi^{\bar i}_+ +  {\bar \epsilon}_+ \psi^{\bar i}_-, \\
\delta\psi^j_- & = &  - {\bar \epsilon}_+ \partial_{z} \phi^i - {\epsilon_-} \Gamma^i_{j k} \psi^j_- \psi^k_{+}, \\
\delta\psi^{\bar j}_+ & = & - {{\epsilon_-}}\partial_{\bar z} \phi^{\bar i} - {\bar \epsilon}_+ \Gamma^{\bar i}_{\bar j \bar k} \psi^{\bar j}_+ \psi^{\bar k}_{-},\\
\delta \psi^i_{+} & = & - {\bar \epsilon}_- \partial_{\bar z} \phi^i - {\epsilon_+} \Gamma^i_{j k} \psi^j \psi^k_{\bar z},\\
\delta \psi^{\bar i}_{-}& = & - {{\epsilon_+}}\partial_z \phi^{\bar i} - {\bar \epsilon}_- \Gamma^{\bar i}_{\bar j \bar k} \psi^{\bar j} \psi^{\bar k}_{z},\
\label{last}
\end{eqnarray}
where $\epsilon_+$, $\epsilon_-$, ${\bar \epsilon}_-$ and ${\bar \epsilon}_+$ are the infinitesimal fermionic parameters associated with the supersymmetries generated by the four supercharges of the $N=(2,2)$ algebra $Q_-$, $Q_+$, ${\overline Q}_+$ and ${\overline Q}_-$ respectively.

A useful point to note at this juncture is that one can obtain the $N=(2,2)$ superalgebra in two-dimensions via a dimensional reduction of the $N=1$ superalgebra in four-dimensions. Consequently, one can obtain (\ref{first})-(\ref{last}) via a dimensional reduction of the supersymmetric field variations that leave an $N=1$ supersymmetric non-linear sigma model in four-dimensions invariant. In turn, by setting $\epsilon_-$ and ${\bar \epsilon}_+$ to zero in (\ref{first})-(\ref{last}),\footnote{Upon twisting, the supersymmetry parameters must now be interpreted as different sections of different line bundles. This is to ensure that the resulting field transformations will remain physically consistent. In particular, the parameters $\epsilon _-$ and ${\bar \epsilon}_+$, associated with the supercharges $Q_+$ and ${\overline Q}_-$, are now sections of the non-trivial bundles ${\overline K}^{-1/2}$ and $K^{-1/2}$ respectively. On the other hand, the parameters $\epsilon_+$ and ${\bar \epsilon}_-$, associated with the supercharges $Q_-$ and ${\overline Q}_+$, are functions on $\Sigma$. One can therefore pick $\epsilon_+$, ${\bar \epsilon}_-$ to be constants, and $\epsilon _-$, ${\bar \epsilon}_+$ to vanish, so that the twisted theory has a global fermionic symmetry generated by the scalar supercharge $Q = Q_- + {\overline Q}_+$, where $Q_- \equiv Q_L$ and ${\overline Q}_+ \equiv Q_R$, as required.} and making the replacements $\psi^j_- \to \psi^j$, $\psi^{\bar j}_+ \to \psi^{\bar j}$, $\psi^j_+ \to \psi^j_{\bar z}$ and $\psi^{\bar j}_- \to \psi^{\bar j}_z$, (which, together are equivalent to twisting the $N=(2,2)$ model into the A-model), we will be able to obtain the field variations in (\ref{varfirst})-(\ref{varlast}) as required (after using the equations of motion $H^i_{\bar z} = \partial_{\bar z}\phi^i$ and $H^{\bar i}_z = \partial_z \phi^{\bar i}$). In short, for one to obtain the explicit field variations generated by the BRST supercharge of the twisted theory, one can  start off with the field variations of the $N=1$ sigma model in four-dimensions, dimensionally reduce them in two dimensions, set the appropriate infinitesimal supersymmetry parameters to zero, and finally redefine the spins of the relevant fields accordingly. This observation will be useful when we discuss the construction of the gauged half-twisted model in section 3.

\newsection{The Half-Twisted Gauged Sigma Model}

We shall now proceed to couple the A-model to a $\it{non}$-$\it{dynamical}$ gauge field which takes values in the Lie algebra spanned by the vector fields generating the associated free $G$-action on $X$. Thereafter, we will discuss the pertinent features of the resulting model which will be most relevant to the later sections of our paper.

\newsubsection{Description of the $G$-Action on $X$}

Let us now suppose that the K\"ahler manifold $X$ admits a compact, $d$-dimensional isometry group $G$, that is, $G$ is a compact group of automorphisms of $X$ which leave fixed its metric and almost complex structure. The infinitesimal generators of this group are given by a set of vector fields on $X$, which, we shall write as $V_a$ for $a = 1, \dots, d$ ($d$ being the dimension of $G$). In other words, the free  $G$-action on $X$ is generated by the vector fields $V_a$.

These fields obey the following conditions. Firstly, they are holomorphic vector fields, which means that their holomorphic (anti-holomorphic) components are holomorphic (anti-holomorphic) functions, that is,
\be
{{\partial V^i_a} \over {\partial \phi^{\bar j}}} ={ {\partial V^{\bar i}_a} \over {\partial \phi^{j}} } = 0.
\label{v1}
\ee
(Note that $V_a = \sum_{i =1}^{n} V_a^i (\partial / \partial \phi^i) + \sum_{\bar i =1}^{n}V_a^{\bar i}( \partial / \partial \phi^{\bar i}$) in component form, where $n =\textrm{dim}_{\mathbb C} X$).

Secondly, the assertion that the $G$-action on $X$ generated by the vector fields $V_a$ for $a = 1, \dots, d$ leave fixed its metric, is equivalent to the assertion that they obey the Killing vector equations
\be
D_i V_{ja} + D_j V_{ia} = 0, \quad D_i V_{\bar j a} + D_{\bar j} V_{ia} = 0,
\label{v2}
\ee
where $D_j$ and $D_{\bar j}$ denote covariant derivatives with respect to the Levi-Civita connection on $X$, while $V_{ia} = g_{i \bar j}V^{\bar j}_a$ and $V_{\bar j a} = g_{i \bar j} V^i_a$.

Finally, the statement that the Killing vector fields $V_a$ generate a $G$-action on $X$ implies that they realise a $d$-dimensional Lie algebra $\mathfrak g$ of $G$, that is, they obey
\be
[V_a, V_b] = f_{ab}{}^c V_c,
\label{v3}
\ee
where $f_{ab}{}^c$ are the structure constants of $G$. One can explicitly write this in component form as
\begin{eqnarray}
[V_a, V_b]^i & = & V^j_a ({\partial V_b^i \over \partial \phi^j}) - V^j_b ({\partial V_a^i \over \partial \phi^j}) \nonumber \\
& = & f_{ab}{}^c V_c^i,\
\label{v4}
\end{eqnarray}
and
\begin{eqnarray}
[V_a, V_b]^{\bar i} & = & V^{\bar j}_a ({\partial V_b^{\bar i} \over \partial \phi^{\bar j}}) - V^{\bar j}_b ({\partial V_a^{\bar i} \over \partial \phi^{\bar j}}) \nonumber \\
& = & f_{ab}{}^c V_c^{\bar i},\
\label{v5}
\end{eqnarray}

\newsubsection{Gauging by the Group $G$}

Note that we want to gauge the half-twisted supersymmetric sigma model by the $d$-dimensional group $G$. What this means geometrically can be explained as follows. Consider the space of maps $\Phi : \Sigma \to X$, which can be viewed as the space of sections of a trivial bundle $M= X \times \Sigma$. If however, one redefines $M$ to be a non-trivial bundle given by $X \hookrightarrow M \to \Sigma$, then $\Phi$ will define a section of the bundle $M$. In other words, $\phi^i(z, \bar z)$ will not represent a map $\Sigma \to X$, but rather, it will be a section of $M$. Thus, since the $\phi^i$'s are no longer functions but sections of a non-trivial bundle, their derivatives will be replaced by covariant derivatives. By introducing a connection on $M$ with $G$ as the structure group, we are actually introducing on $\Sigma$ gauge fields $A^a$, which, locally, can be regarded as $G$-valued one-forms with the usual gauge transformation law ${A^a}' = g^{-1} A^a g + g^{-1} dg$, whereby $g \in G$. This is equivalent to gauging the sigma model by $G$.

\newsubsection{Constructing the Half-Twisted Gauged Sigma Model}

In order to gauge the half-twisted supersymmetric sigma model by the $d$-dimensional group $G$, one will need to introduce, in the formulation,  $d$  gauge multiplets, each  consisting of the two-dimensional gauge field $A^a$, its fermionic gaugino superpartner $\psi^a$, and the complex scalar $\phi^a$, with values in the Lie algebra $\mathfrak g$ and transforming in the adjoint representation of $G$. These fields will appear as the components of the two-dimensional vector superfields ${\cal V}^a$ of $N=(2,2)$ superspace, where each ${\cal V}^a$  can be expanded as
\begin{eqnarray}
{\cal V}^a & = & \theta^- {\bar \theta}^- A^a_{z} + \theta^+ {\bar \theta}^+ A^a_{\bar z} - \theta^- {\bar \theta}^+\phi^a - \theta^+ {\bar \theta}^- {\bar \phi}^a + i \theta^- \theta^+ ( {\bar \theta}^-{\bar \psi}^a_- + {\bar \theta}^+ {\bar \psi}^a_+ ) + i {\bar \theta}^+ {\bar \theta}^- ( \theta^- \psi^a_- +  \theta^+ \psi^a_+ ) \nonumber \\
&&  + \theta^- \theta^+  {\bar \theta}^+ {\bar \theta}^- D^a.\
\label{calV}
\end{eqnarray}
Here, the $\theta$'s are the anticommuting coordinates of $N=(2,2)$ superspace, and the $D^a$'s are real, auxillary scalar superfields which can be eliminated from the final Lagrangian via the relevant equations of motion. Also, on $\Sigma$, the gauge fields $A^a_z$ and $A^a_{\bar z}$ can be regarded as connection $(1,0)$- and $(0,1)$-forms, the $\phi^a$'s and ${\bar \phi}^a$'s can be regarded as complex scalars, while the (${\psi}^a_+$, ${\bar \psi}^a_+$)'s, and (${\psi}^a_-$, ${\bar \psi}^a_-$)'s can be regarded as worldsheet spinors given by sections of the bundles ${\overline K}^{1/2}$ and $K^{1/2}$  respectively.

Since our aim is to construct a half-twisted gauged sigma model, we must also twist the above fields of the gauge multiplet, as we had done so with the fields $\phi^i$, $\phi^{\bar i}$, $\psi^i_+$, $\psi^{\bar i}_+$, $\psi^i_-$ and $\psi^{\bar i}_-$ of the $N=(2,2)$ sigma model to arrive at the A-model. Recall from the footnote on pg. 7 that in an A-twist, the spin of each field will be redefined as $s' = s + {q_V \over 2}$, where $s$ is its original spin, and $q_V$ is its corresponding vector R-charge. Hence, in order to ascertain how the fields of the gauge multiplet can be A-twisted, we must first determine their vector R-charges. To this end, note that a vector R-rotation is effected by the transformations $\theta^{\pm} \to e^{-i \alpha} \theta^{\pm}$ and ${\bar \theta}^{\pm} \to e^{i \alpha} {\bar \theta}^{\pm}$, where $\alpha$ is a real parameter of the rotation. Equivalently, one can see from (\ref{calV}), that under a vector R-rotation, the fields of the gauge multiplet will transform as $(A^a_z, A^a_{\bar z}, \phi^a, {\bar \phi}^a) \to (A^a_z, A^a_{\bar z}, \phi^a, {\bar \phi}^a)$, $\psi^a_{\pm} \to e^{i \alpha} \psi^a_{\pm}$, and ${\bar \psi}^a_{\pm} \to e^{-i \alpha} {\bar \psi}^a_{\pm}$. In other words, the fields $(A^a_z, A^a_{\bar z}, \phi^a, {\bar \phi}^a)$ have $q_V = 0$, the $\psi^a_{\pm}$ have $q_V = 1$, and the ${\bar \psi}^a_{\pm}$ have $q_V = -1$. This means that under an A-twist, $A^a_z$ and $A^a_{\bar z}$ will remain as connection $(1,0)$- and $(0,1)$-forms on $\Sigma$, while $\phi^a$ and ${\bar \phi}^a$ will remain as complex scalars. However, ${\bar \psi}^a_-$ and $\psi^a_+$ will now be complex scalars, while $\psi^a_-$ and ${\bar \psi}^a_+$ are $(1,0)$- and $(0,1)$-forms on $\Sigma$ respectively. For clarity, we shall re-label (${\bar \psi}^a_-$, $\psi^a_+$) as (${\bar \psi}^a$, $\psi^a$), and ($\psi^a_-$, ${\bar \psi}^a_+$) as ($\psi^a_z$, ${\bar \psi}^a_{\bar z}$), in accordance with their properties on $\Sigma$.

Next,  let us determine the generalisation of (\ref{varfirst})-(\ref{varlast}) in the presence of the gauge multiplet of fields. To this end, we can extend the recipe outlined at the end of section 2.4 to the gauged case. Essentially, one can begin by considering the supersymmetric field transformations which leave an $N=1$, gauged non-linear sigma model invariant (see pg. 50 of \cite{bagger}), dimensionally reduce them in two dimensions, and set the supersymmetry parameters $\epsilon_-$ and ${\bar \epsilon}_+$ to zero. In doing so, we obtain the generalisation of (\ref{varfirst})-(\ref{varlast}) as
\begin{eqnarray}
\label{gauge tx first}
\delta \phi^a & = & 0, \\
\delta\phi^i & = & \epsilon_+ \psi^i, \\
\delta \phi^{\bar i} & = & {\bar \epsilon}_- \psi^{\bar i},\\
\delta A^a_z & = & {\bar \epsilon}_- \psi^a_z, \\
\delta A^a_{\bar z} & = & \epsilon_+ \psi^a_{\bar z}, \\
\delta\psi^j & = & - i {\bar \epsilon}_- \phi^a V_a^j, \\
\delta\psi^{\bar j} & = & - i {\epsilon}_+ \phi^a V_a^{\bar j}, \\
\delta\psi^a_z & = & - i \epsilon_+ D_z \phi^a, \\
\delta \psi^a_{\bar z} & = & - i {\bar \epsilon}_- D_{\bar z} \phi^a, \\
\delta \psi^i_{\bar z} & = & - {\bar \epsilon}_- H^i_{\bar z} - {\epsilon_+} \Gamma^i_{j k} \psi^j \psi^k_{\bar z}, \\
\delta \psi^{\bar i}_{z} & = & - {\epsilon_+} H^{\bar i}_{z} - {\bar \epsilon_-} \Gamma^{\bar i}_{\bar j \bar k} \psi^{\bar j} \psi^{\bar k}_z, \
\label{gauge tx last}
\end{eqnarray}
where one recalls that $\epsilon_+$ and ${\bar \epsilon}_-$ are the constant parameters associated with the scalar BRST supercharges $Q_L$ and $Q_R$ respectively. $D_z$ and $D_{\bar z}$ are the covariant derivatives with respect to the connection one-forms $A^a_z$ and $A^a_{\bar z}$ respectively.\footnote{One can explicitly write $D_z \phi^a = \partial_z \phi^a + f^a{}_{bc} A^a_z \phi^c$ and $D_{\bar z} \phi^a = \partial_{\bar z} \phi^a + f^a{}_{bc} A^b_{\bar z} \phi^c$.}  In order to determine how the auxillary fields $H^i_{\bar z}$ and $H^{\bar i}_z$ should transform, one just needs to insist that the field transformations generated by $Q=Q_L + Q_R$ are nilpotent up to a gauge transformation. In particular, we must have (after setting $\epsilon_+$ and ${\bar \epsilon}_-$ to 1 for notational simplicity)
\be
\delta^2 \psi^i_{\bar z} = - i \phi^a (\partial_k V_a^i) \psi^k_{\bar z}
\ee
and
\be
\delta^2 \psi^{\bar i}_z = - i \phi^a (\partial_{\bar k} V_a^{\bar i}) \psi^{\bar k}_{z},
\ee
which then means that we must have
\be
\delta H^i_{\bar z} =  R^i{}_{k \bar j l}\psi^k \psi^{\bar j} \psi^l_{\bar z} + i \phi^a (D_j V^i_a) \psi^j_{\bar z}    -    \Gamma^i_{jk} \psi^j H^k_{\bar z}
\label{varhi}
\ee
and
\be
\delta H^{\bar i}_{z}  =   R^{\bar i}{}_{\bar j l \bar k}\psi^{\bar j}\psi^l \psi^{\bar k}_{z} + i \phi^a (D_{\bar j} V^{\bar i}_a) \psi^{\bar j}_{z}  -  \Gamma^{\bar i}_{\bar j \bar k} \psi^{\bar j} H^{\bar k}_{z}.
\label{varhbari}
\ee
Notice that since $\epsilon_+$ and ${\bar \epsilon}_-$ are constants, the fields on the LHS and RHS of (\ref{gauge tx first})-(\ref{gauge tx last}) have the same worldsheet spins; the twist of the gauge multiplet fields is consistent with the field transformations (\ref{gauge tx first})-(\ref{gauge tx last}) as expected. Furthermore, one finds from (\ref{gauge tx first})-(\ref{gauge tx last}) that
\begin{eqnarray}
\delta^2 \phi^j  =  -i \phi^a V^j_a, & \quad & \delta^2 \phi^{\bar j}  =  -i \phi^a V^{\bar j}_a, \\
\delta^2 A^a_z = -i D_z\phi^a, & \quad & \delta^2 A^a_{\bar z} = -i D_{\bar z} \phi^a, \\
\delta^2 \psi^a_z = -i f^a{}_{bc} \psi^b_z \phi^c, & \quad & \delta^2 \psi^a_z = -i f^a{}_{bc} \psi^b_z \phi^c,\\
\delta^2 \psi^j  =  -i \phi^a (\partial_k V^j_a) \psi^k, & \quad & \delta^2 \psi^{\bar j}  =  -i \phi^a (\partial_{\bar k} V^{\bar j}_a) \psi^{\bar k}\
\end{eqnarray}
and $\delta^2 \phi^a = 0$, as required of a gauged model. Hence, we are now ready to define our gauge- and BRST-invariant Lagrangian by generalising the results of section 2.1.

To obtain a gauge-invariant generalisation of $S$ in (\ref{S}), we will need to obtain a gauge-invariant generalisation of (\ref{Stop}). This can be achieved by replacing the partial derivatives in $V$ of (\ref{V}), with gauge covariant derivatives. Moreover, in doing so, we only introduce terms which do not modify the overall ghost number. This means that we will be able to retain a classical ghost number symmetry as desired. Note also that we only want to couple the sigma model to a non-dynamical gauge multiplet of fields. In other words, we will not include a super-field-strength term for the gauge multiplet in defining the action. Therefore, the action of our half-twisted gauged sigma model can be written as
\be
S_{\textrm{gauged}} = \int_{\Sigma}|d^2 z| \{Q, V_{\textrm{gauged}} \},
\label{Sgauged}
\ee
where
\be
V_{\textrm{gauged}} = i g_{i \bar j} ( \psi^i_{\bar z} D_z \phi^{\bar j} + \psi^{\bar j}_z D_{\bar z} \phi^i - {1\over 2} \psi^{\bar j}_z H^i_{\bar z} - {1\over 2} \psi^i_{\bar z}H^{\bar j}_z),
\label{Vgauged}
\ee
such that from the field transformations in (\ref{gauge tx first})-(\ref{gauge tx last}) and (\ref{varhi})-(\ref{varhbari}), we find that
\begin{eqnarray}
S_{\textrm{gauged}} & = & \int_{\Sigma} |d^2z| \ ( g_{i\bar j} D_{\bar z} \phi^i D_z \phi^{\bar j} + g_{i \bar j} \psi^i_{\bar z} {\widehat D}_z \psi^{\bar j} + g_{i \bar j} \psi^{\bar j}_z {\widehat D}_{\bar z} \psi^i + g_{i \bar j} \psi^i_{\bar z} \psi^a_z V^{\bar j}_a + g_{i \bar j} \psi^{\bar j}_z \psi^a_{\bar z} V^i_a \nonumber  \\
&& \hspace{0.5cm}  -{i \over 2} g_{i \bar j}\psi^{\bar j}_z \psi^j_{\bar z} (D_j V^i_a) \phi^a -{i \over 2} g_{i \bar j}\psi^{i}_{\bar z} \psi^{\bar k}_{z} (D_{\bar k} V^{\bar j}_a) \phi^a +g_{i \bar j} \psi^i_{\bar z} \Gamma^{\bar j}_{\bar l \bar k}A^a_z V^{\bar l}_a \psi^{\bar k}  \nonumber  \\
&& \hspace{0.5cm} +g_{i \bar j} \psi^{\bar j}_{z} \Gamma^{i}_{l k}A^a_{\bar z} V^{l}_a \psi^{k} - R_{\bar m k \bar j l}\psi^{\bar m}_z \psi^k \psi^{\bar j} \psi^l_{\bar z}).\
\label{Sgauged explicit}
\end{eqnarray}
Note that we have used the equations of motion $H^{\bar i}_z = D_z \phi^{\bar i}$ and $H^i_{\bar z} = D_{\bar z} \phi^i$ to eliminate the auxillary fields $H^{\bar i}_z$ and $H^i_{\bar z}$ in our computation of $S_{\textrm{gauged}}$ above. Notice also that as desired, there are no kinetic terms for the non-dynamical fields $A^a_z$, $A^a_{\bar z}$, $\psi^a_z$, $\psi^a_{\bar z}$ and $\phi^a$ in $S_{\textrm{gauged}}$. However, the various covariant derivatives in $S_{\textrm{gauged}}$ are now given by
\begin{eqnarray}
D_{\bar z} \phi^{i} & = & \partial_{\bar z} \phi^i + A^a_{\bar z} V^i_a, \\
D_{z} \phi^{\bar j} & = & \partial_{z} \phi^{\bar j} + A^a_{z} V^{\bar j}_a, \\
D_j V^i_a & = & \partial_{j} V^i_a  + \Gamma^i_{jl} V^l_a, \\
D_{\bar k} V^{\bar j}_a & = & \partial_{\bar k} V^{\bar j}_a  + \Gamma^{\bar j}_{\bar k \bar l} V^{\bar l}_a, \\
{\widehat D}_z \psi^{\bar j} & = & \partial_z \psi^{\bar j} + A^a_z \partial_{\bar k} V^{\bar j}_a \psi^{\bar k} + \partial_z \phi^{\bar i} \Gamma^{\bar j}_{\bar i \bar l} \psi^{\bar l}, \\
{\widehat D}_{\bar z} \psi^{i} & = & \partial_{\bar z} \psi^{i} + A^a_{\bar z} \partial_{j} V^{i}_a \psi^{j} + \partial_{\bar z} \phi^{j} \Gamma^{i}_{j k} \psi^{k}. \
\end{eqnarray}
Under the classical ghost number symmetry of (\ref{Sgauged explicit}), we find that the fields $\psi^i$, $\psi^{\bar i}_{ z}$, $\psi^{\bar i}$ and $\psi^i_{\bar z}$ can be assigned the $(g_L, g_R)$ left-right ghost numbers $(1,0)$, $(-1,0)$, $(0,1)$ and $(0,-1)$ respectively as in the ungauged model, while the fields of the gauge multiplet $A^a_z$, $A^a_{\bar z}$, $\psi^a_z$, $\psi^a_{\bar z}$ and $\phi^a$ can be assigned the $(g_L, g_R)$ left-right ghost numbers $(0,0)$, $(0,0)$, $(0,1)$, $(1,0)$ and $(1,1)$.

\newsubsection{Ghost Number Anomaly}

As a relevant digression, let us now discuss the ghost number anomaly of the half-twisted gauged sigma model. In this paper, we are considering the case where $G$ is  unitary and abelian. As we will see in section 4, this means that $\partial_i V^j_a = \partial_{\bar i} V^{\bar j}_a = 0$. Consequently, $S_{\textrm{gauged}}$ can be simplified to
\begin{eqnarray}
S'_{\textrm{gauged}} & = & \int_{\Sigma} |d^2z| \ ( g_{i\bar j} D_{\bar z} \phi^i D_z \phi^{\bar j} + g_{i \bar j} \psi^i_{\bar z} {D}_z \psi^{\bar j} + g_{i \bar j} \psi^{\bar j}_z {D}_{\bar z} \psi^i + g_{i \bar j} \psi^i_{\bar z} \psi^a_z V^{\bar j}_a + g_{i \bar j} \psi^{\bar j}_z \psi^a_{\bar z} V^i_a \nonumber  \\
&& \hspace{0.5cm}  -{i \over 2} g_{i \bar j}\psi^{\bar j}_z \psi^j_{\bar z} \Gamma^i_{jl} V^l_a \phi^a -{i \over 2} g_{i \bar j}\psi^{i}_{\bar z} \psi^{\bar k}_{z} \Gamma^{\bar j}_{\bar k \bar l} V^{\bar l}_a \phi^a +g_{i \bar j} \psi^i_{\bar z} \Gamma^{\bar j}_{\bar l \bar k}A^a_z V^{\bar l}_a \psi^{\bar k}  \nonumber  \\
&& \hspace{0.5cm} +g_{i \bar j} \psi^{\bar j}_{z} \Gamma^{i}_{l k}A^a_{\bar z} V^{l}_a \psi^{k} - R_{\bar m k \bar j l}\psi^{\bar m}_z \psi^k \psi^{\bar j} \psi^l_{\bar z}).\
\label{S'gauged explicit}
\end{eqnarray}
In general, the non-minimally coupled terms in $S'_{\textrm{gauged}}$ which are not part of any covariant derivative but involve the non-dynamical fields, do not affect anomalies. This is because anomalies are by definition what cannot be eliminated by any choice of regularisation, and in a particular choice such as the Pauli-Villars  scheme, one regularises by adding higher order derivatives to the kinetic energy, which can be taken to be independent of these auxillary fields even if they appear in the classical action $S'_{\textrm{gauged}}$.\footnote{The author wishes to thank Ed Witten for helpful email correspondences on this point.} In addition, note that in sigma model peturbation theory, the four-fermi term $R_{\bar m k \bar j l}\psi^{\bar m}_z \psi^k \psi^{\bar j} \psi^l_{\bar z}$ can be treated as a perturbation which does not affect the computation of the anomaly either (just as in the case with the  A-model with action (\ref{S})). Since the $\phi^i$ and $\phi^{\bar i}$ fields have vanishing ghost numbers, the ghost number anomaly  can then be calculated via the index theorem associated with the $D_z$ and $D_{\bar z}$ operators acting on $\psi^{\bar j}$ and $\psi^i$, which are sections of the pullback bundles $\Phi^*(\overline {TX})$ and $\Phi^*(TX)$ respectively. Notice that we have the same considerations as in the A-model. Hence, via similar arguments to that in sect. 2.3 on the non-vanishing of correlation functions of dimension $(0,0)$ operators, one must have the condition
\be
\left( \int_{\Sigma}\Phi^*c_1(TX) + \textrm{dim}_{\mathbb C} X(1-g) \right)   > 0,
\label{indexgauged}
\ee
where $g$ is the genus of the worldsheet Riemann surface $\Sigma$. Note that one will be considering degree-zero maps $\Phi$ in the perturbative limit. Therefore, from (\ref{indexgauged}), it is clear that  for the half-twisted gauged sigma model in perturbation theory, the relevant worldsheet will also be a genus-zero, simply-connected Riemann surface.

\newsubsection{Important Features of the Half-Twisted Gauged Sigma Model}

We shall now explore some important features of the half-twisted gauged sigma model with action $S_{\textrm{gauged}}$ given in (\ref{Sgauged explicit}).  Classically, the trace of the stress tensor from $S_{\textrm{gauged}}$ vanishes, i.e., $T_{\bar z z} = 0$. The other non-zero components of the stress tensor are given by
\be
T_{zz} = g_{i \bar j} \partial_z \phi^i (\partial_z \phi^{\bar j} + A^a_z V^{\bar j}_a) + g_{i \bar j} \psi^{\bar j}_z  \left(\partial_z\psi^i + \Gamma^i_{jk} \partial_z \phi^j \psi^k \right)
\label{gauge tzz}
\ee
and
\be
T_{\bar z \bar z} =g_{i \bar j}  (\partial_{\bar z} \phi^i + A^a_{\bar z}V_a^i) \partial_{\bar z} \phi^{\bar j} + g_{i \bar j}  \psi_{\bar z}^i  ( \partial_{\bar z} \psi^{\bar j} + \Gamma^{\bar j}_{\bar l \bar k}\partial_{\bar z} \phi^{\bar l} \psi^{\bar k}).
\label{gauge tbarzbarz}
\ee
Furthermore, one can go on to show that
\be
T_{\bar z \bar z} = \{ Q_R , i g_{i \bar j} \psi_{\bar z}^i \partial_{\bar z} \phi^{\bar j} \},
\label{gauge tbarzbarz short}
\ee
and
\be
T_{zz} = \{ Q_L , i g_{i \bar j} \psi_{z}^{\bar j} \partial_{z} \phi^{i} \}.
\label{gauge tzz short}
\ee
In addition, we also have
\begin{eqnarray}
\label{QRtzz}
[Q_R , T_{zz} ] & = & - {1\over 2}\ g_{i \bar j} \psi^{\bar j}_z  \left(\partial_z\phi^k (D_k V_a^i) \phi^a + 2 \partial_z \phi^a V_a^i \right) \nonumber \\
& \neq & 0 \hspace{0.2cm} (\textrm{even on-shell}).
\end{eqnarray}
Before we proceed further, recall that that the operators and states of the half-twisted gauged sigma model are in the $Q_R$-cohomology. Note also that $Q^2_L = Q^2_R = 0$, even though $Q^2 = 0$ up to a gauge transformation only. Next, from (\ref{gauge tbarzbarz short}), we see that $T_{\bar z \bar z}$ is $Q_R$-exact (and thus $Q_R$-invariant) and therefore trivial in $Q_R$-cohomology. Also, from (\ref{QRtzz}), we see that $T_{zz}$ is not in the $Q_R$-cohomology. Consequently, one can make the following observations about the half-twisted gauged sigma model.

\vspace{0.4cm}{\noindent{\it Spectrum of Operators and Correlation Functions}}

Firstly, since $T_{ \bar z z} = 0$, the variation of the correlation functions due to a change in the scale of $\Sigma$ will be given by $\left <{{\cal O}_1(z_1)} {{\cal O}_2(z_2)} \dots {{\cal O}_s(z_s)} T_{\bar z z} \right >= 0$. In other words, the correlation functions of local physical operators will continue to be invariant under arbitrary scalings of $\Sigma$. Thus, the correlation functions are always independent of the K\"ahler structure on $\Sigma$ and may depend only on its complex structure.\footnote{However, as will be shown in section 4, the correlation functions of the subset of operators that are also in the $Q_L$-cohomology, will be independent of the metric and complex structure of $\Sigma$ and even $X$.} In addition, $T_{zz}$ is holomorphic in $z$; from the conservation of the stress tensor, we have $\partial_{\bar z}T_{zz} = - \partial_z T_{\bar z z} = 0$.

Secondly, note  that the $\partial_{\bar z}$ operator on $\Sigma$ is given by ${\bar L}_{-1} = \oint d{\bar z}\ T_{\bar z \bar z}$.  This means that  ${\partial_{\bar z}\left <{\cal O}_1(z_1) {\cal O}_2(z_2) \dots {\cal O}_s(z_s)  \right >}$ will be given by $ \oint d{\bar z} \left <T_{\bar z \bar z} \ {\cal O}_1(z_1) {\cal O}_2(z_2) \dots {\cal O}_s(z_s) \right >$. This vanishes because   $T_{\bar z \bar z} = \{ Q_R, \dots \}$ and therefore, $T_{\bar z \bar z} \sim 0$ in $Q_R$-cohomology. Thus, the correlation functions of local operators are always holomorphic in $z$. Likewise, we can also show that $\cal O$, as an element of the $Q_R$-cohomology, varies homolomorphically with $z$. Indeed, since the momentum operator (which acts on $\cal O$ as $\partial_{\bar z}$) is given by $\bar L_{-1}$, the term $\partial_{\bar z} \cal O$ will be given by the commutator $[ \bar L_{-1}, \cal O]$. Since $\bar L_{-1} = \oint d\bar z\,T_{\bar z\bar z}$, we will  have $\bar L_{-1}=\{{Q_R},V_{-1}\}$ for some $V_{-1}$. Hence, because $\cal O$ is physical such that ${\{Q_R, \cal O\}} =  0$, it will be true that $\partial_{\bar z}{\cal O}=\{{Q_R},[V_{-1},{\cal O}]\}$ and thus vanishes in $Q_R$-cohomology.

We can make a third and important observation as follows. But first, note that we say that a local operator $\cal O$ inserted at the origin has
dimension $(n,m)$ if under a rescaling $z\to \lambda z$, $\bar z\to \bar\lambda z$, it transforms as $\partial^{n+m}/\partial z^n\partial\bar z^m$, that is, as $\lambda^{-n}\bar\lambda{}^{-m}$. Classical local operators have dimensions $(n,m)$ where $n$ and $m$ are non-negative integers. However, only local operators with $m = 0$ survive in $Q_R$-cohomology. The reason for the last statement is that the rescaling of $\bar z$ is generated by $\bar L_0=\oint d\bar z\, \bar z T_{\bar z\bar z}$.  As we saw above, $T_{\bar z\,\bar z}$ is of the form $\{Q_R,\dots\}$, so $\bar L_0=\{{Q}_R,V_0\} $ for some $V_0$. If $\cal O$ is to be admissible as a local physical operator, it must at least be true that $\{{Q}_R, {\cal O}\}=0$. Consequently, $[\bar L_0,{\cal O}]=\{{Q}_R,[V_0,{\cal O}]\}$.  Since the eigenvalue of $\bar L_0$ on $\cal O$ is $m$, we have $[\bar L_0,{\cal O}]=m{\cal O}$. Therefore, if $m\not= 0$, it follows that ${\cal O}$ is $Q_R$-exact and thus trivial in $Q_R$-cohomology. A useful fact to note at this point is that via the same arguments, since $T_{zz}$ is of the form $\{Q_L, \dots \}$, only operators with $n=0$ survive in $Q_L$-cohomology. These two facts will be important in section 4.

Also, from the last paragraph, we have the condition $\bar L_0 = 0$ for operators in the $Q_R$-cohomology. Let the spin of any operator be $S$, where $S= L_0 - \bar L_0$. Since after twisting, $Q_R$ is a scalar BRST operator of spin zero, we will have  $[S, Q_R]= 0$. This in turn implies that $[Q_R, L_0 ] = 0$. In other words, the operators of the half-twisted gauged sigma model will remain in the $Q_R$-cohomology after global dilatations of the worldsheet coordinates.

Last but not least, note that the coefficients of the mode expansion of $T_{zz}$ generate arbitrary holomorphic reparameterisations of $z$. Hence, since $T_{zz}$ is not $Q_R$-closed, the operators will not remain in the $Q_R$-cohomology after arbitrary holomorphic reparameterisations of coordinates on $\Sigma$. This also means that $\oint dz [Q_R, T_{zz} ] = [Q_R, L_{-1} ] \neq 0$.\footnote{Since we are working modulo $Q_R$-trivial operators, it suffices for $T_{zz}$ to be holomorphic up to $Q_R$-trival terms before an expansion in terms Laurent coefficients is permitted.}   Therefore, the operators will not remain in the $Q_R$-cohomology after global translations on the worldsheet.

Note that these observations are based on the fact that $T_{\bar z z}$, $T_{\bar z \bar z}$ or $T_{zz}$ either vanishes or is absent in $Q_R$-cohomology. In perturbation theory, where quantum effects are small enough, cohomology classes can only be destroyed and not created. Thus, if it is true classically that a cohomology either vanishes or is absent, it should continue to be true at the quantum level. Hence, the above observations will hold in the quantum theory as well.

\vspace{0.4cm}{\noindent{\it  A Holomorphic Chiral Algebra $\cal A$}}

Let ${\cal O} (z)$ and $\widetilde {\cal O} (z')$ be two $Q_R$-closed operators such that their product is $Q_R$-closed as well. Now, consider their operator product expansion or OPE:
\be
{\cal O}(z)  {\widetilde {\cal O}}(z') \sim \sum_k f_k (z-z') {\cal O}_k (z'),
\label{OPE}
\ee
in which the explicit form of the coefficients $f_k$ must be such that the scaling dimensions and $(g_L, g_R)$ ghost numbers of the operators agree on both sides of the OPE. In general, $f_k$ is not holomorphic in $z$. However, if we work modulo $Q_R$-exact operators in passing to the $Q_R$-cohomology, the $f_k$'s which are non-holomorphic and are thus not annihilated by $\partial / \partial {\bar z}$, drop out from the OPE because they multiply operators ${\cal O}_k$ which are $Q_R$-exact. This is true because $\partial / \partial{\bar z}$ acts on the LHS of (\ref{OPE}) to give terms which are cohomologically trivial.\footnote{Since $\{Q_R,{\cal O}\}=0$, we have $\partial_{\bar z}{\cal O}=\{Q_R, V(z)\}$ for some $V(z)$, as argued before. Hence $\partial_{\bar z}{\cal O}(z)\cdot {\widetilde {\cal O}}(z')=\{Q_R,V(z){\widetilde {\cal O}}(z')\}$.} In other words, we can take the $f_k$'s to be holomorphic coefficients in studying the $Q_R$-cohomology. Thus, the OPE of (\ref{OPE}) has a holomorphic structure. Hence, we have established that the $Q_R$-cohomology of holomorphic local operators has a natural structure of a holomorphic chiral algebra (in the sense that the operators obey (\ref{OPE}), and are annihilated by only one of the two scalar BRST generators $Q_R$ of the supersymmetry algebra) which we shall denote as $\cal A$.

\vspace{0.4cm}{\noindent{\it The Important Features of $\cal A$}}

In summary, we have established that $\cal A$ is always preserved under global dilatations and Weyl scalings, though (unlike the usual physical notion of a chiral algebra) it is not preserved under general holomorphic coordinate transformations and global translations on the Riemann surface $\Sigma$ (since $T_{zz}$ is not in the $Q_R$-cohomology even at the classical level). Likewise, the OPEs of the chiral algebra of local operators obey the usual relations of holomorphy, associativity, invariance under dilatations of $z$, and Weyl scalings, but not invariance under arbitrary holomorphic reparameterisations and global translations of $z$.\footnote{However, as will be shown in section 4.3, the correlation functions of the subset of operators in $\cal A$ that are also in the $Q_L$-cohomology, will be topological invariants of $\Sigma$ and even $X$.} The local operators are of dimension (n,0) for $n \geq 0$, and the chiral algebra of such operators requires a flat metric up to scaling on $\Sigma$ to be defined.\footnote{Notice that we have implicitly assumed the flat metric on $\Sigma$ in all of our analysis thus far.} Therefore, the chiral algebra that we have obtained can either be globally-defined on a Riemann surface of genus one, or be locally-defined on an arbitrary but curved $\Sigma$. We shall assume the latter in this paper. Finally, as is familiar for chiral algebras, the correlation functions of these operators may depend on $\Sigma$ only via its complex structure. The correlation functions are holomorphic in the parameters of the theory and are therefore protected from perturbative corrections.

\newsection{The Relation to the Chiral Equivariant Cohomology}

We will now proceed to demonstrate the connection between the half-twisted gauged sigma model in perturbation theory and the chiral equivariant cohomology. To this end, we shall specialise to the case where the gauge group $G$ is abelian. As a result of our analysis, some of the established mathematical results on the chiral equivariant cohomology can be shown to either lend themselves to straightforward physical explanations, or be verified through purely physical reasoning. Moreover, one can also determine fully, the de Rham cohomology ring of $X/G$, from a topological chiral ring generated by the local   ground operators of the chiral algebra $\cal A$.

\newsubsection{The Half-Twisted Abelian Sigma Model at Weak Coupling}

We shall start by discussing the theory in the limit of weak coupling or infinite-volume of $X$. We will then proceed to show that the desired results hold at all values of the coupling constant and hence, to all orders in perturbation theory, in the final  subsection. But firstly, by an expansion of the Lagrangian in $S_{\textrm{gauged}}$ of (\ref{Sgauged explicit}), we have

\begin{eqnarray}
{\cal L}_{\textrm{gauged}} & = & g_{i \bar j} \partial_{\bar z} \phi^i \partial_z \phi^{\bar j} + g_{i \bar j} \partial_{\bar z} \phi^i A^a_z V_a^{\bar j} + g_{i \bar j} \partial_z \phi^{\bar j} A^a_{\bar z} V^i_a + g_{i \bar j} A^a_{\bar z} V_a^i A^b_z V^{\bar j}_b  \nonumber \\
&& + \psi_{\bar z \bar j} \partial_z \psi^{\bar j} + \psi_{\bar z \bar j}\partial_z \phi^{\bar i} \Gamma^{\bar j}_{\bar i \bar l} \psi^{\bar l} + \psi_{\bar z \bar j} A^a_z \partial_{\bar k} V^{\bar j}_a \psi^{\bar k} + \psi_{z i} \partial_{\bar z} \psi^i  \nonumber \\
&&  + \psi_{z i} \partial_{\bar z} \phi^j \Gamma^i_{j k} \psi^k   +  \psi_{zi} A^a_{\bar z} \partial_j V^i_a \psi^j  + \psi_{\bar z \bar j} \Gamma^{\bar j}_{\bar l \bar k}A^a_z V^{\bar l}_a \psi^{\bar k} + \psi_{z i} \Gamma^i_{l k} A^a_{\bar z} V^l_a \psi^k  \nonumber \\
&& + \psi_{\bar z \bar j} \psi^a_z V^{\bar j}_a  + \psi_{z i} \psi^a_{\bar z} V^i_a - {i\over 2} g^{j \bar m} \psi_{z i} \psi_{\bar z \bar m} ( \partial_j V^i_a + \Gamma^i_{jk} V^k_a) \phi^a \nonumber \\
&& - {i \over 2} g^{\bar k m} \psi_{\bar z \bar j} \psi_{z m} ( \partial_{\bar k} V^{\bar j}_a  + \Gamma^{\bar j}_{\bar k \bar n}V^{\bar n}_a ) \phi^a  - g^{\bar m n} g^{l \bar n}R_{\bar m  k \bar j l}\psi_{z n} \psi^k \psi^{\bar j} \psi_{\bar z \bar n}, \
\label{lgauged}
\end{eqnarray}
where we have rewritten $g_{i \bar j} \psi^{\bar j}_z$ as $\psi_{zi}$, and $g_{i \bar j} \psi^i_{\bar z}$ as $\psi_{\bar z \bar j}$. Next, recall from (\ref{v4})-(\ref{v5})  that we have the relations
\begin{eqnarray}
[V_a , V_b]^i & = &   V^j_a \partial_j V^i_b - V^j_b \partial_j V^i_a \nonumber  \\
&& f_{ab}{}^c V_c^i  \
\label{4}
\end{eqnarray}
and
\begin{eqnarray}
[V_a , V_b]^{\bar i} & = & V^{\bar j}_a \partial_{\bar j} V^{\bar i}_b - V^{\bar j}_b \partial_{\bar j} V^{\bar i}_a \nonumber \\
&& f_{ab}{}^c V_c^{\bar i}.  \
\label{5}
\end{eqnarray}
If we consider $G$ to be a unitary, abelian gauge group such as $U(1)^d = T^d$ for any $d \geq 1$, then the structure constants $f_{ab}{}^c$ must vanish for all $a, b, c = 1, 2, \dots, d$, that is, $[V_a, V_b]^i = [V_a, V_b]^{\bar i} = 0$. Since the generators of the $U(1)$'s are unique,      that is, $V_a \neq V_b \neq 0$, from (\ref{4})-({\ref {5}), it will mean that $\partial_j V^i_a = \partial_{\bar j} V^{\bar i}_a = 0$ for  abelian $G = T^d$.  Hence, ${\cal L}_{\textrm{gauged}}$ can be simplified to
\begin{eqnarray}
{\cal L}_{\textrm{abelian}} & = & g_{i \bar j} \partial_{\bar z} \phi^i \partial_z \phi^{\bar j} + g_{i \bar j} \partial_{\bar z} \phi^i A^a_z V_a^{\bar j} + g_{i \bar j} \partial_z \phi^{\bar j} A^a_{\bar z} V^i_a + g_{i \bar j} A^a_{\bar z} V_a^i A^b_z V^{\bar j}_b  \nonumber \\
&& + \psi_{\bar z \bar j} \partial_z \psi^{\bar j} + \psi_{\bar z \bar j}\partial_z \phi^{\bar i} \Gamma^{\bar j}_{\bar i \bar l} \psi^{\bar l} + \psi_{z i} \partial_{\bar z} \psi^i + \psi_{z i} \partial_{\bar z} \phi^j \Gamma^i_{j k} \psi^k \nonumber \\
&&  + \psi_{\bar z \bar j} \Gamma^{\bar j}_{\bar l \bar k}A^a_z V^{\bar l}_a \psi^{\bar k} + \psi_{z i} \Gamma^i_{l k} A^a_{\bar z} V^l_a \psi^k  + \psi_{\bar z \bar j} \psi^a_z V^{\bar j}_a  + \psi_{z i} \psi^a_{\bar z} V^i_a \nonumber \\
&&  - {i\over 2} g^{j \bar m} \psi_{z i} \psi_{\bar z \bar m} \Gamma^i_{jk} V^k_a \phi^a  - {i \over 2} g^{\bar k m} \psi_{\bar z \bar j} \psi_{z m} \Gamma^{\bar j}_{\bar k \bar n}V^{\bar n}_a  \phi^a  \nonumber \\
&& - g^{\bar m n} g^{l \bar n}R_{\bar m  k \bar j l}\psi_{z n} \psi^k \psi^{\bar j} \psi_{\bar z \bar n}. \
\label{labelian}
\end{eqnarray}

Now consider the action
\begin{eqnarray}
{\cal L}_{\textrm{equiv}} & = &  p_{z i} \partial_{\bar z}\phi^i  + p_{\bar z \bar j} \partial_z \phi^{\bar j} + \psi_{z i} \partial_{\bar z} \psi^i + \psi_{\bar z \bar j} \partial_z \psi^{\bar j}  - g^{\bar j i} (p_{z i} - \Gamma^k_{i l} \psi_{z k} \psi^l)(p_{\bar z \bar j} - \Gamma^{\bar k}_{\bar j \bar l}\psi_{\bar z \bar k} \psi^{\bar l}) \nonumber \\
&&  - g^{\bar m n} g^{l \bar n}R_{\bar m  k \bar j l}\psi_{z n}\psi_{\bar z \bar n} \psi^k \psi^{\bar j} + g_{i \bar j} \partial_{\bar z} \phi^i A^a_z V_a^{\bar j} + g_{i \bar j} \partial_z \phi^{\bar j} A^a_{\bar z} V^i_a + g_{i \bar j} A^a_{\bar z} V_a^i A^b_z V^{\bar j}_b \nonumber \\
&&  + \psi_{\bar z \bar j} \Gamma^{\bar j}_{\bar l \bar k}A^a_z V^{\bar l}_a \psi^{\bar k} + \psi_{z i} \Gamma^i_{l k} A^a_{\bar z} V^l_a \psi^k  + \psi_{\bar z \bar j} \psi^a_z V^{\bar j}_a  + \psi_{z i} \psi^a_{\bar z} V^i_a \nonumber \\
&&  - {i\over 2} g^{j \bar m} \psi_{z i} \psi_{\bar z \bar m} \Gamma^i_{jk} V^k_a \phi^a  - {i \over 2} g^{\bar k m} \psi_{\bar z \bar j} \psi_{z m} \Gamma^{\bar j}_{\bar k \bar n}V^{\bar n}_a  \phi^a.  \
\label{lequiv}
\end{eqnarray}
From ${\cal L}_{\textrm{equiv}}$ above, the equations of motion for the fields $p_{zi}$ and $p_{\bar z \bar j}$ are given by
\be
p_{z i} =  g_{i \bar j} \partial_z \phi^{\bar j} + \Gamma^k_{il} \psi_{zk} \psi^l \quad \textrm{and}\quad  p_{\bar z \bar j} = g_{i \bar j} \partial_{\bar z} \phi^i + \Gamma^{\bar k}_{\bar j \bar l} \psi_{\bar z \bar k}\psi^{\bar l}.
\label{pi}
\ee
By substituting the above explicit expressions of $p_{zi}$ and $p_{\bar z \bar j}$ back into (\ref{lequiv}), one obtains ${\cal L}_{\textrm{abelian}}$. In other words, ${\cal L}_{\textrm{abelian}}$ and ${\cal L}_{\textrm{equiv}}$ define the same theory. Hence, we shall take ${\cal L}_{\textrm{equiv}}$ to be the Lagrangian of the half-twisted abelian sigma model instead of ${\cal L}_{\textrm{abelian}}$. The reason for doing so is that we want to study the sigma model in the weak-coupling regime where the coupling tends to zero, or equivalently, the infinite-volume limit. For this purpose, ${\cal L}_{\textrm{equiv}}$ will soon prove to be more useful.

Before we proceed to consider the infinite-volume limit, we shall discuss a further simplification of ${\cal L}_{\textrm{equiv}}$. Now recall that the two-dimensional gauge field $A$ defines a connection one-form on some vector bundle over the Riemann surface $\Sigma$. Let the curvature two-form of the bundle be $F$. Since $\Sigma$ is of complex dimension one, it will means that the $(2,0)$ and $(0,2)$ components of the curvature two-form $F_{zz}$ and $F_{\bar z \bar z}$ respectively, must be zero. Since we shall be considering the worldsheet $\Sigma$ to be a simply-connected, genus-zero Riemann surface in perturbation theory, we can consequentially write the corresponding holomorphic and anti-holomorphic components of the connection one-form $A$ in pure gauge, that is,
\be
A_z = i \partial_z (U^{\dagger})^{-1} \cdot U^{\dagger}
\label{Az}
\ee
and
\be
A_{\bar z} = i \partial_{\bar z} U \cdot U^{-1},
\label{Abarz}
\ee
where $U \in G$. Equations (\ref{Az}) and (\ref{Abarz}) show that either $A_z$ or $A_{\bar z}$ may be set to zero by a gauge transformation, but in general not simultaneously. However, since we considering $U$ to be abelian and unitary, or rather, $U^{\dagger} = U^{-1}$, we can set both $A_z$ and $A_{\bar z}$ to zero in ${\cal L}_{\textrm{equiv}}$ \cite{GSW2}. In addition, from varying the fields $\psi^a_z$ and $\psi^a_{\bar z}$ in ${\cal L}_{\textrm{equiv}}$, we have the equations of motion $\psi_{\bar z \bar j} V^{\bar j}_a = \psi_{zi} V^i_a = 0$. Hence, ${\cal L}_{\textrm{equiv}}$ can be further simplified to
\begin{eqnarray}
{\cal L}_{\textrm{equiv}'} & = &  p_{zi} \partial_{\bar z}\phi^i  + p_{\bar z \bar j} \partial_z \phi^{\bar j} + \psi_{z i} \partial_{\bar z} \psi^i + \psi_{\bar z \bar j} \partial_z \psi^{\bar j}  - g^{i\bar j} (p_{zi} - \Gamma^k_{i l} \psi_{z k} \psi^l)(p_{\bar z \bar j} - \Gamma^{\bar k}_{\bar j \bar l}\psi_{\bar z \bar k} \psi^{\bar l}) \nonumber \\
&&  - g^{\bar m n} g^{l \bar n}R_{\bar m  k \bar j l}\psi_{z n}\psi_{\bar z \bar n} \psi^k \psi^{\bar j}  - {i\over 2} g^{i \bar j} ( \psi_{z l} \psi_{\bar z \bar j} \Gamma^l_{ik} V^k_a \phi^a  + \psi_{\bar z \bar l} \psi_{z i} \Gamma^{\bar l}_{\bar j \bar n}V^{\bar n}_a  \phi^a).  \
\label{l'equiv}
\end{eqnarray}

Finally, we consider the infinite-volume or weak-coupling limit, whereby $g_{i \bar j} \to \infty$ or the inverse metric $g^{i \bar j} \to 0$. In this limit, ${\cal L}_{\textrm{equiv}'}$ will read as
\be
{\cal L}_{\textrm{weak}}  =   p_{zi} \partial_{\bar z}\phi^i  + p_{\bar z \bar j} \partial_z \phi^{\bar j} + \psi_{z i} \partial_{\bar z} \psi^i + \psi_{\bar z \bar j} \partial_z \psi^{\bar j}.
\label{lweak}
\ee
Thus, one can regard ${\cal L}_{\textrm{weak}}$ as the effective Lagrangian of the weakly-coupled, half-twisted gauged sigma model with unitary, abelian gauge group $G=U(1)^d$ for any $d \geq 1$.

From the equations of motion associated with ${\cal L}_{\textrm{weak}}$, we find that $\partial_{\bar z} \phi^i$, $\partial_{z} \phi^{\bar i}$, $\partial_{\bar z} p_{zi}$,  $\partial_{z} p_{\bar z \bar i}$, $\partial_{\bar z} \psi^i$, $\partial_{z} \psi^{\bar i}$, $\partial_{\bar z} \psi_{z i}$ and $\partial_{z} \psi_{\bar z \bar i}$ must vanish, that is, the fields are solely dependent on either $z$ or $\bar z$ accordingly. In addition, via standard field theory methods, we find from ${\cal L}_{\textrm{weak}}$ the following OPE's
\be
p_{zi}(z) \phi^j( w) \sim - {\delta^j_i \over {z-w}}, \qquad \psi_{zi} (z) \psi^j( w) \sim  {\delta^j_i \over {z-w}},
\label{OPE1}
\ee
and
\be
p_{\bar z \bar i} (\bar z) \phi^{\bar j}( \bar w) \sim -{\delta^{\bar j}_{\bar i} \over {\bar z- \bar w}}, \qquad \psi_{\bar z \bar i}(\bar z) \psi^{\bar j}( \bar w) \sim {\delta^{\bar j}_{\bar i} \over {\bar z- \bar w}}.
\label{OPE2}
\ee
Notice that (\ref{OPE1}) and (\ref{OPE2}) are the usual OPE's of the conformal $bc$-$\beta\gamma$ system and its complex conjugate respectively; the fields $p_{zi}$, $\phi^j$, $\psi_{zi}$, $\psi^j$, $p_{\bar z \bar i}$, $\phi^{\bar j}$, $\psi_{\bar z \bar i}$ and $\psi^{\bar j}$, correspond to the fields $\beta_i$, $\gamma^j$, $b_i$, $c^j$, ${\bar \beta}_{\bar i}$, ${\bar \gamma}^{\bar j}$, ${\bar b}_{\bar i}$ and ${\bar c}^{\bar j}$. In other words, ${\cal L}_{\textrm{weak}}$ defines a conformal system which is a tensor product of a $bc$-$\beta\gamma$ system and its complex conjugate.

\newsubsection{The Spectrum of Operators and the Chiral Equivariant Cohomology}

\vspace{0.4cm}{\noindent{\it The Fock Vacuum}}

Note that since the fields $p_{zi}$, $\phi^i$, $\psi_{zi}$, $\psi^i$, $p_{\bar z \bar i}$, $\phi^{\bar i}$, $\psi_{\bar z \bar i}$, $\psi^{\bar i}$ are solely dependent on either $z$ or $\bar z$, we can express them in terms of a Laurent expansion. And since the fields $p_{zi}$, $\psi_{zi}$, $p_{\bar z \bar i}$,  $\psi_{\bar z \bar i}$ scale as dimension one fields, while $\phi^i$, $\psi^i$, $\phi^{\bar i}$, $\psi^{\bar i}$ scale as dimension zero fields, their corresponding Laurent expansions will be given by
\be
p_{zi} = \sum_{n \in {\mathbb Z}} {{p_{i,n}} \over {z^{n+1}}}, \qquad \qquad p_{\bar z \bar i} = \sum_{n \in {\mathbb Z}} {{p_{\bar i,n}} \over {{\bar z}^{n+1}}},
\label{piexpand}
\ee
\be
\psi_{z i} = \sum_{n \in {\mathbb Z}} {{\psi_{i,n}} \over {z^{n+1}}}, \qquad \qquad \psi_{\bar z \bar i} = \sum_{n \in {\mathbb Z}} {{\psi_{\bar i,n}} \over {{\bar z}^{n+1}}},
\label{psizexpand}
\ee
\be
\phi^i = \sum_{n \in {\mathbb Z}} {{\phi^i_n} \over {z^n}}, \qquad \qquad  \phi^{\bar i} = \sum_{n \in {\mathbb Z}} {{\phi^{\bar i}_n} \over {{\bar z}^n}},
\label{phiexpand}
\ee
and
\be
\psi^i = \sum_{n \in {\mathbb Z}} {{\psi^i_n} \over {z^n}}, \qquad \qquad \psi^{\bar i} = \sum_{n \in {\mathbb Z}} {{\psi^{\bar i}_n} \over {{\bar z}^n}}.
\label{psiexpand}
\ee
In addition, from the OPE's in (\ref{OPE1})-(\ref{OPE2}), we find that their mode expansion coefficients obey the relations
\be
[\phi^i_n, p_{j,m}] = \delta^i_j \delta_{n, -m}, \qquad \qquad \{\psi^i_n, \psi_{j, m} \} = \delta^i_j \delta_{n , -m},
\label{commutation1}
\ee
and
\be
[\phi^{\bar i}_n, p_{\bar j, m}] = \delta^{\bar i}_{\bar j} \delta_{n, -m}, \qquad \qquad \{\psi^{\bar i}_n, \psi_{\bar j, m} \} = \delta^{\bar i}_{\bar j} \delta_{n , -m},
\label{commutation2}
\ee
with all other commutation and anti-commutation relations between fields vanishing. Consequently, from (\ref{commutation1}) and (\ref{commutation2}) above, we find that the zero modes obey
\be
[p'_{j,0}, \phi^i_0] = \delta^i_j, \qquad [\phi^{\bar i}_0, p_{\bar j, 0} ] = \delta^{\bar i}_{\bar j},
\label{zeromodes1}
\ee
and
\be
\{ \psi_{j,0}, \psi^i_0 \} = \delta^i_j, \qquad \{\psi^{\bar i}_0, \psi_{\bar j, 0} \} = \delta^{\bar i}_{\bar j}.
\label{zeromodes2}
\ee
where we have rewritten $- p_{j,m}$ as $p'_{j,m}$ for convenience.

Notice that (\ref{zeromodes1}) and (\ref{zeromodes2}) are identical to the relations $[a, a^{\dagger}] = 1$ and $\{a, a^{\dagger}\} = 1$ between the annihilation and creation operators $a$ and $a^{\dagger}$ respectively; $p'_{j,0}$, $\phi^{\bar i}_0$, $\psi_{j,0}$ and $\psi^{\bar i}_0$ will correspond to annihilation operators while $\phi^i_0$, $p_{\bar j, 0}$, $\psi^i_0$ and $\psi_{\bar j, 0}$ will correspond to creation operators. Next, let us denote the Fock vacuum for the zero mode sector of the Hilbert space of states by $|0 \rangle$. Then one has the condition that
\be
 p'_{j,0} |0\rangle  = \phi^{\bar i}_0 |0\rangle = \psi_{j,0} |0\rangle = \psi^{\bar i}_0 |0\rangle = 0.
\label{vacuum1}
\ee
Recall that in the state-operator correspondence, $|0\rangle$ is represented by the identity operator. Therefore,  (\ref{vacuum1}) implies that the corresponding vertex operators of the theory must be independent of the fields $\phi^{\bar i}$, $\psi^{\bar i}$ and their derivatives.\footnote{In general, the vertex operators need not be independent of the derivatives of the fields $\phi^{\bar i}$ and $\psi^{\bar i}$. However, recall from section 3.5 that in the half-twisted gauged sigma model, the operators must have scaling dimension $(n,0)$ for $n \geq 0$. This means that the they must be independent of the $\bar z$-derivatives of the fields $\phi^{\bar i}$ and $\psi^{\bar i}$. In addition, we have the condition $\partial_z \phi^{\bar i} = \partial_z \psi^{\bar i} = 0$. Hence, the operators must be independent of any worldsheet derivatives of $\phi^{\bar i}$ and $\psi^{\bar i}$ to any non-zero order.} However, because $p_{zi}$ and $\psi_{z i}$ are of (holomorphic) weight one, we can still consider these fields and their $z$-derivatives (but not their $\bar z$-derivatives since they are holomorphic in $z$) in the corresponding operator expressions.\footnote{From the Laurent expansion of the dimension $(1,0)$ fields $p_{zi}$ and $\psi_{zi}$, we find that unless $\psi_{j, -1} |0\rangle$ or $p'_{j,-1} |0 \rangle$ is zero, we may still include them in the corresponding vertex operator expressions.}

\vspace{0.4cm}{\noindent{\it Physical Operators and the Sheaf of CDR on $X$}}

From the various discussions so far, we learn that the physical operators in the $Q_R$-cohomology must comprise only of the fields $p_{zi}$, $\phi^i$, $\psi_{zi}$, $\psi^i$, $\phi^a$, $\psi^a_z$ and their $z$-derivatives of order greater or equal to one. (Recall from section 3.5 that the operators of the half-twisted gauged sigma model must be of scaling dimension $(n,0)$ where $n \geq 0$ only, so they cannot consist of $p_{\bar z \bar i}$, $\psi^a_{\bar z}$ and the $\bar z$-derivatives of any field.) As explained in section 3.5, these physical operators in the chiral algebra $\cal A$ must be locally-defined over $\Sigma$. However, they remain globally-defined over $X$. Hence, from the OPE's in (\ref{OPE1}), and the corresponding mode relations in (\ref{commutation1}), we find that they will correspond to global sections of the sheaf $\Omega^{ch}_X \otimes \langle \psi^a_z, \phi^a \rangle$, where $\Omega^{ch}_X$ is the chiral de Rham complex on $X$ \cite{MSV1}, and $\langle \psi^a_z, \phi^a \rangle$ is a free polynomial algebra generated by the commuting and non-commuting operators $\partial^k_z \phi^a$ and $\partial^k_z \psi^a_{z}$, where $k \geq 0$. Note also that $\langle \psi^a_z, \phi^a \rangle$ is a polynomial algebra that is symmetric in $\partial^k_z \phi^a$ and antisymmetric in $\partial^k_z \psi^a_{z}$.

Now, let $V_a = \sum_{i=1}^{\textrm{dim}_{\mathbb C}X}V^i_a (\partial/ \partial {\phi^i})$ be a holomorphic vector field on $X$ which generates a $G$-action, such that the holomorphic components $V^i_a$ realise a subset of the corresponding Lie algebra $\mathfrak g$ of $G$. As in \cite{MC,CDO}, one can proceed to define a   dimension one operator $J_{V_a} (z) =  p_{zi}V^i_a (z)$ of ghost number zero, where its conformally-invariant and hence conserved charge $K_{V_a} = \oint J_{V_a} dz$ will generate a local symmetry of the two-dimensional theory on $\Sigma$. From the first OPE in (\ref{OPE1}), we find that
\be
J_{V_a}(z) \phi^k(z')\sim -{V^k_a (z' )\over z-z'}.
\label{jV}
\ee
Under the symmetry transformation generated by $K_{V_a}$, we have $\delta \phi^k = i \epsilon [ K_{V_a}, \phi^k ]$. Thus, we see  from (\ref{jV}) that $K_{V_a}$ generates an infinitesimal  holomorphic diffeomorphism $\delta\phi^k= -i \epsilon V^k_a$ associated with the $G$-action on the target space $X$. For finite diffeomorphisms, we have a general field transformation ${\tilde \phi}^k = g^k (\phi^i)$ induced by the $G$-action on $X$, where each $g^k (\phi^i)$ is a holomorphic function in the $\phi^i$s.  In addition, one can also compute that
\be
J_{V_a}(z) p_{zk}(z') \sim {{p_{zi} \partial_k V^i_a (z' )}\over {z-z'}}.
\label{jV2}
\ee
However, since we are considering the case where $G=T^d$ is unitary and abelian, the right-hand side of (\ref{jV2}) vanishes, as a trivial structure constant implies that $\partial_k V^i_a = 0$. Hence, the OPE of $J_{V_a}$ with $d(p_{zi})$, an arbitrary polynomial function in $p_{zi}$ and its $z$-derivatives, is trivial.

Next, consider adding to $J_{V_a}$ another ghost number zero dimension one operator, consisting of the fermionic fields, given by $J_F (z) = \psi^n t_n{}^m \psi_{zm} (z)$, where $t[\phi]$ is some matrix holomorphic in the $\phi^i$'s,  with the indices $n,m = 1, \dots, \textrm{dim}_{\mathbb C}X$. Once again, its conformally-invariant and hence conserved charge $K_F = \oint J_F dz$ will generate a local symmetry of the two-dimensional theory on $\Sigma$. From the OPE's in (\ref{OPE1}), we find that
\be
J_F(z) \psi^n(z') \sim {  {\psi^m (z') t_m{}^n  }  \over z-z'},
\label{jF1}
\ee
while
\be
J_F(z) \psi_{z n} (z') \sim - {  { t_n{}^m \psi_{z, m} (z')  }  \over z-z'}.
\label{jF2}
\ee
Under the symmetry transformation generated by $K_F$, we have $\delta \psi^n = i \epsilon [ K_F, \psi^n]$ and $\delta \psi_{z n} = i \epsilon [ K_F, \psi_{z n}]$. Hence, we see from (\ref{jF1}) and (\ref{jF2}) that $K_F$ generates the infinitesimal transformations $\delta \psi^n=i \epsilon \psi^m t_m{}^n$ and $\delta \psi_{z n}= - i \epsilon t_n{}^m \psi_{z m}$. For finite transformations, we will have ${\tilde \psi}^n = \psi^m A_m{}^n$ and ${\tilde \psi}_{z n} = (A^{-1})_n{}^m \psi_{zm}$, where $[A(\phi)]$ is a matrix holomorphic in the $\phi^i$'s given by $[A(\phi)] = e^{i \alpha [t(\phi)]}$, where $\alpha$ is a finite transformation parameter.  Recall at this point that the $\psi^n$'s transform as holomorphic sections of the pull-back $\Phi^*(TX)$, while the $\psi_{z n}$'s transform as holomorphic sections of the pull-back $\Phi^* (T^*X)$. Moreover, note that the transition function matrix of a dual bundle is simply the inverse of the transition function matrix of the original bundle. Hence, this means that if we are using an appropriate symmetry of the worldsheet theory (and hence $[t(\phi)]$) to `glue' their local descriptions over an arbitrary intersection $U_1 \cap U_2$, we can consistently identify $[A(\phi)]$ as the holomorphic transition matrix of the tangent bundle $TX$. (This was was done in \cite{MC} to derive the automorphism relations of the sheaf of CDR defined in \cite{MSV1}). However, this need not be the case in general, and for $K_F$ to still generate a symmetry of the worldsheet theory, it is sufficient that $[A(\phi)]$ and therefore $[t(\phi)]$ be arbitrary matrices which are holomorphic in the $\phi^i$'s.

For the purpose of connecting with the results in \cite{andy1,andy2} by Lian et al., let $t_m{}^n (z) = {\partial V^n/ \partial \phi^m}$. Thus, the total dimension one current operator $J_{V_a} + J_F$, with charges $K_{L} = K_{V_a} + K_F$ generating the symmetries discussed above, will be given by (after rewriting $p_{zi}$, $\phi^j$, $\psi_{zi}$, $\psi^j$  as $\beta_i$, $\gamma^j$, $b_i$, $c^j$)
\be
L_{V_a} (z) = \beta_i V^i_a (z) + {{\partial V^j_a} \over{\partial \gamma^i}}c^i b_j (z),
\label{LV}
\ee
where the normal ordering symbol has been omitted for notational simplicity. As defined in section 3 of \cite{andy1}, the dimension (or conformal weight) one operator $L_{V_a}(z)$ is just a vertex algebraic analogue of the Lie derivative with respect to the holomorphic vector field $V_a$ on $X$. Indeed, one can compute the OPE
\be
L_{V_a} (z) L_{V_b} (z') \sim { L_{[V_a, V_b]} (z') \over {z-z'}},
\label{LLOPE}
\ee
which is a vertex algebraic analogue of the differential-geometric relation between two Lie derivatives $[ L_{\xi}, L_{\eta}] = L_{[\xi, \eta]}$, where $\xi$ and $\eta$ are any two vector fields on $X$. Note that the operator observables of our gauge-invariant model ought to be $G$-invariant, where one recalls that $G$ is the compact gauge group of automorphisms on $X$; an admissible operator $\cal O$ will be invariant under the field transformations induced by the $G$-action. In other words, we will have $[K_{L}, {\cal O}\} = 0$, where $K_{L}$ is the conserved charge generating the field transformations associated with the $G$-action. This means that the operator product expansion $L_{V_a}(z) {\cal O}(z')$ should not contain any single poles. However, because we are considering the case where $G=T^d$ is unitary and abelian, we have a further simplification of $L_{V_a}(z)$; the second term on the right-hand side of $L_{V_a}(z)$ vanishes since $\partial V^j_a / \partial \gamma^i = 0$. Hence, $L_{V_a}(z)$ effectively acts as $J_{V_a}(z)=  p_{zi} V^i_a(z)$ on the $Q_R$-cohomology of operators in the abelian theory. Since a general, local operator ${\cal O}$ must comprise only of the fields $p_{zi}$, $\phi^i$, $\psi_{zi}$, $\psi^i$, $\phi^a$, $\psi^a_z$ and their $z$-derivatives, it can be expressed as $f(\phi^i) d(p_{zi}) g(\psi^i, \psi_{zi})s(\phi^a, \psi^a_z)$, where $g(\psi^i, \psi_{zi})$ is a polynomial function up to some finite order in $\psi^i$, $\psi_{zi}$ and their $z$-derivatives (since $\psi^i$ and $\psi_{zi}$ are anti-commuting Grassmannian fields), while $s(\phi^a, \psi^a_z)$ is a polynomial function in $\phi^a$, $\psi^a_z$ and their $z$-derivatives up to some finite order in $\partial^k_z\psi^a_z$ for $k\geq 0$ (since $\psi^a_z$ is an anti-commuting Grassmannian field). Note that the operator product expansions of $J_{V_a}$ with the fields $p_{zi}$, $\psi^i$, $\psi_{zi}$, $\phi^a$ and $\psi^a_z$ are non-singular, and since the operator product expansion $J_{V_a} (z) {\cal O} (z')$ cannot contain single poles, we deduce that the operator product expansion $L_{V_a} (z)f(z')$ cannot contain single poles either, that is, $[K_{L}, {f}(z)] = 0$. In other words, for $\cal O$ to be an admissible operator in the abelian theory, it would suffice that $f(\phi^i)$ be a $G$-invariant holomorphic function in $\phi^i$. However, by a suitable averaging over the compact group $G$, one can take ${\cal O}  = f(\phi^i) d(p_{zi}) g(\psi^i, \psi_{zi})s(\phi^a, \psi^a_z)$ to be $G$-invariant without changing its cohomology class. Therefore, in either the abelian or non-abelian case, $\cal O$ will be given by a global section of the sheaf $(\Omega^{ch}_X )^{\mathfrak t \geq} \otimes \langle \phi^a, \psi^a_z \rangle$, where $(\Omega^{ch}_X )^{\mathfrak t \geq}$ just denotes the subspace of $\Omega^{ch}_X$ that is invariant under the (worldsheet) symmetry transformation associated with $L_{V_a}(z)$.\footnote{We can always rewrite $(\Omega^{ch}_X  \otimes \langle \phi^a, \psi^a_z \rangle)^{\mathfrak t \geq}$ as $(\Omega^{ch}_X )^{\mathfrak t \geq} \otimes \langle \phi^a, \psi^a_z \rangle$, since the sections of the sheaf $\langle \phi^a, \psi^a_z \rangle$ will always be invariant under the symmetry generated by $K_L$ anyway.}

\vspace{0.4cm}{\noindent{\it About the BRST Operators $Q_L$ and $Q_R$}}

Let us continue by discussing the BRST operators $Q_L$ and $Q_R$ in the regime of weak coupling. To this end, let us first note that the field variations due to $Q_L$ acting on any operator $\cal O$ are
\begin{eqnarray}
\label{deltaL1}
\delta_L \phi^i = \psi^i, & \qquad  \delta_L \psi_{zi} = -p_{zi},  \qquad & \delta_L \psi^a_z = - i \partial_z \phi^a, \\
\label{deltaL2}
\delta_L p_{zi} = 0, & \qquad  \delta_L \psi^i = 0,  \qquad & \delta_L \phi^a = 0. \
\end{eqnarray}
On the other hand, the  non-vanishing field variations due to $Q_R$ acting on any operator $\cal O$ are (after absorbing $i$ via a trivial field redefinition of $\phi^a$)
\begin{eqnarray}
\label{deltaR1}
\delta_R \phi^i = 0, & \qquad  \delta_R \psi_{zi} = 0,  \qquad & \delta_R \psi^a_z = 0, \\
\label{deltaR2}
\delta_R \phi^a = 0, & \qquad  \delta_R \psi^i = -\phi^a V_a^i,  \qquad & \delta_R p_{zi} = 0, \
\end{eqnarray}
where $ \delta_R p_{zi} = 0$ only upon using the appropriate equations of motion.\footnote{By using the equations of motion from ${\cal L}_{\textrm{equiv}'}$, we find that $\delta_R p_{zi} = -{1 \over 2} \psi_{zl} g^{l\bar j}(g_{i \bar j, k} V_a^k + g_{i \bar j, {\bar k}} V_a^{\bar k}) \phi^a$. However, in sigma model perturbation theory, derivatives of the metric are of order $R^{-1}_c$, where $R_c$ is the characteristic radius of curvature of the target space $X$. Thus, in the infinite-volume limit where $R_c \to \infty$, the derivatives of the metric vanish, and $\delta_R p_{zi} = 0$ follows.}

From ${\cal L}_{\textrm{weak}}$, we find that the corresponding supercurrents can be written (where normal ordering is understood) as
\be
Q_L(z) = p_{zi} \psi^i (z) \qquad \textrm{and} \qquad Q_R(z) = - \phi^a V_a^i \psi_{zi}(z),
\label{supercurrents}
\ee
so that
\be
Q_L  = \oint {dz \over {2 \pi i}} \ p_{zi} \psi^i (z)  \qquad \textrm{and} \qquad Q_R = - \oint {dz \over {2 \pi i}} \ \phi^a V_a^i \psi_{zi} (z).
\label{supercharges}
\ee
Note that we have the OPE's
\be
Q_L(z) Q_L(z') \sim \textrm{reg} \qquad \textrm{and} \qquad Q_R(z) Q_R(z') \sim \textrm{reg}.
\ee
Hence, from (\ref{supercharges}), we see that $\{Q_L, Q_L\}$ and $\{Q_R, Q_R\}$ vanish, that is, $Q^2_L = Q^2_R = 0$. Another point to note is that $Q_L$ and $Q_R$ have ghost numbers $(1,0)$ and $(0,1)$ respectively; $Q_L$ acts to increase the left ghost number of any operator by one, while $Q_R$ acts to increase the right ghost number of any operator by one. In addition, one also has the OPE
\begin{eqnarray}
Q_L(z) Q_R (z') & \sim & {{\phi^a L_{V_a} (z')} \over {z- z'}}.\
\end{eqnarray}
This means that we will have
\be
\{Q_L, Q_R\} = Q_{L_V},
\label{QQ}
\ee
where
\be
Q_{L_V} = \oint {{dz} \over {2 \pi i}} J_{L_V} (z),
\label{QLV}
\ee
and $J_{L_V} (z) = \phi^a L_{V_a}(z)$. Since the OPE's of $\phi^a$ and $L_{V_a}$   with   any admissible operator $\cal O$ do not contain any single poles, we deduce that $Q_{L_V}$ annihilates $\cal O$, that is,
\be
[\{Q_L, Q_R\} , {\cal O} \} = 0.
\label{QLQR}
\ee
To illustrate an important consequence of (\ref{QLQR}), let us take ${\cal O}_a$ to be an admissible fermionic operator of ghost number $(q,p-1)$. Then, from (\ref{QLQR}), we have
\be
[Q_L, \{Q_R , {\cal O}_a\} ] + [Q_R, \{ Q_L , {\cal O}_a\}] = 0.
\ee
If $\{Q_L, {\cal O}_a \} = 0$, we will have $[Q_L,  \{Q_R, {\cal O}_a\}] = 0$. This can be trivially satisfied if $\{Q_R, {\cal O}_a \} = 0$. However, if $\{Q_R, {\cal O}_a \} \neq 0$, because ${Q^2_L} = 0$, one can hope to find an operator ${\cal O}'_a$ of ghost number $(q-1, p)$, such that $\{Q_R, {\cal O}_a\} = \{Q_L, {\cal O}'_a\}$. This important observation will be useful below.

\vspace{0.4cm}{\noindent \it {A Spectral Sequence and the Subset of Operators in the $Q_L$-Cohomology}}

Building towards our main objective of uncovering the physical interpretation of the chiral equivariant cohomology, we would now like to study the subset of operators which are also in the $Q_L$-cohomology, that is, the subset of operators which are also closed with respect to $Q_L$ $\it{and}$ $Q_R$, and can neither be written as a (anti)commutator with $Q_L$ nor $Q_R$. Clearly, they wil also be closed with respect to $Q= Q_L + Q_R$. Hence, in order to ascertain this subset of operators, let us first try to determine the operators in the $Q_R$-cohomology which are also $Q$-closed.

As explained in section 3.5, operators in the $Q_R$-cohomology must have scaling dimension $(n,0)$ where $n \geq 0$. Therefore, let us begin with  a general operator, corresponding to a global section of $(\Omega^{ch}_X)^{\mathfrak t \geq}$, of scaling dimension or conformal weight $(0,0)$, which hence may be admissible as a class in the $Q_R$-cohomology:
\be
{\cal O}_A = A_{i_1 i_2 \dots i_n}(\phi^k) \psi^{i_1} \psi^{i_2} \dots \psi^{i_n}.
\label{OA}
\ee
(Note that we have not included the $\phi^a$ field in ${\cal O}_A$ because it will soon appear naturally in our current attempt to determine the  operators which are $Q$-closed.)  Let us denote $\Delta {\cal O}_A$ as the change in ${\cal O}_A$ due to the action of $Q$, that is,
\be
\Delta {\cal O}_A = \{Q_L, {\cal O}_A \} + \{ Q_R, {\cal O}_A \}.
\label{deltaOA}
\ee
Let us choose ${\cal O}_A$ such that it can be annihilated by $Q_L$, that is, $\{Q_L, {\cal O}_A\} = 0$, so that it may be admissible as a class in the $Q_L$-cohomology as well. Then,
\begin{eqnarray}
\Delta {\cal O}_A & = & \{ Q_R, {\cal O}_A \} \nonumber \\
&=& -i n \phi^a V_a^{i_1} A_{i_1 i_2 \dots i_n}  \psi^{i_2} \dots \psi^{i_n}. \
\end{eqnarray}
Thus, we find that ${\cal O}_A$ is neither in the $Q_R$-cohomology nor    $Q$-closed as required. These observations suggest that corrections to the operator ${\cal O}_A$ need to be made. To this end, recall from our discussion above on   ${\cal O}_a$,    that since ${\cal O}_A$ is to be    admissible as an operator     and is $Q_L$-closed, we may have
\begin{eqnarray}
\{Q_R, {\cal O}_A\} & = & - \{Q_L, {\cal O}^1_A\}\nonumber \\
& = & -i n \phi^a V_a^{i_1}  A_{i_1 i_2 \dots i_n} \psi^{i_2} \dots \psi^{i_n}, \
\label{spectral1}
\end{eqnarray}
where ${\cal O}^1_A$ is a global section of the sheaf $(\Omega^{ch}_X)^{\mathfrak t \geq} \otimes \langle \phi^a, \psi^a_z \rangle$. One may then `refine' the definition of ${\cal O}_A$ to
\begin{eqnarray}
{\widehat {\cal O}}_A  & =  & {\cal O}_A + {\cal O}^1_A \nonumber \\
&=& A_{i_1 i_2 \dots i_n} (\phi^k) \psi^{i_1} \psi^{i_2} \dots \psi^{i_n} + \phi^a A_{a i_1 i_2 \dots i_{n-2}}(\phi^k) \psi^{i_1} \psi^{i_2} \dots \psi^{i_{n-2}}, \
\label{OA1}
\end{eqnarray}
where
\be
\partial_m A_{a i_1 i_2 \dots i_{n-2}}  \psi^m \psi^{i_1}\psi^{i_2} \dots \psi^{i_{n-2}} = n V_a^{i_1} A_{i_1 i_2 i_3 \dots i_n} \psi^{i_2} \psi^{i_3} \dots \psi^{i_n}.
\label{c1}
\ee
Then, the change in ${\widehat{\cal O}}_A$ due to the action of $Q$ will be given by
\begin{eqnarray}
\Delta {\widehat{\cal O}}_A & = & \{Q_R, {\cal O}^1_A \} \nonumber \\
&=& -i (n-2)  \phi^a \phi^b V_b^{i_1} A_{a i_1 i_2 i_3 \dots i_{n-2}}\psi^{i_2} \psi^{i_3} \dots \psi^{i_{n-2}}. \
\end{eqnarray}
Notice that $\Delta{\widehat{\cal O}}_A$ is two orders lower in the fermionic fields $\psi^i$'s than $\Delta {\cal O}_A$. This indicates that if we continue to refine ${\widehat{\cal O}}_A$ in the above fashion, we will eventually reach $\Delta {\widehat{\cal O}}_A = 0$, and obtain the exact expression of the $Q$-closed operator as desired. To verify this statement, let us continue to refine ${\widehat{\cal O}}_A$ by adding to it another term ${\cal O}^2_A$, that is,
\begin{eqnarray}
{\widehat {\cal O}}_A  & =  & {\cal O}_A + {\cal O}^1_A + {\cal O}^2_A \nonumber \\
&=& A_{i_1 i_2 \dots i_n} (\phi^k) \psi^{i_1} \psi^{i_2} \dots \psi^{i_n} + \phi^a A_{a i_1 i_2 \dots i_{n-2}}(\phi^k) \psi^{i_1} \psi^{i_2} \dots \psi^{i_{n-2}} \nonumber \\
&& + \phi^a \phi^b A_{a b i_1 i_2 \dots i_{n-4}}(\phi^k) \psi^{i_1} \psi^{i_2} \dots \psi^{i_{n-4}}, \
\label{OA2}
\end{eqnarray}
whereby
\begin{eqnarray}
\{Q_R, {\cal O}^1_A\} & = & - \{Q_L, {\cal O}^2_A\}\nonumber \\
& = &  -i (n-2)  \phi^a \phi^b V_b^{i_1} A_{a i_1 i_2 i_3 \dots i_{n-2}}\psi^{i_1} \psi^{i_3} \dots \psi^{i_{n-2}}, \
\label{spectral2}
\end{eqnarray}
and therefore
\be
\partial_m A_{a b i_1 i_2 \dots i_{n-4}}  \psi^m \psi^{i_1}\psi^{i_2} \dots \psi^{i_{n-4}} = (n-2) V_b^{i_1} A_{a i_1 i_2 i_3 \dots i_{n-2}} \psi^{i_2} \psi^{i_3}  \dots \psi^{i_{n-2}}.
\label{c2}
\ee
So now, we have
\begin{eqnarray}
\Delta {\widehat{\cal O}}_A & = & \{Q_R, {\cal O}^2_A \} \nonumber \\
&=& -i (n-4)  \phi^a \phi^b \phi^c V_c^{i_1} A_{a b i_1 i_2 i_3 \dots i_{n-4}}\psi^{i_2} \psi^{i_3} \dots \psi^{i_{n-4}}. \
\end{eqnarray}
Indeed, if we continue with the above refining process, we will eventually obtain the correct expression for ${\widehat{\cal O}}_A$ that is $Q$-closed:
\begin{eqnarray}
{\widehat {\cal O}}_A &=& A_{i_1 i_2 \dots i_n} (\phi^k) \psi^{i_1} \psi^{i_2} \dots \psi^{i_n} + \phi^a A_{a i_1 i_2 \dots i_{n-2}}(\phi^k) \psi^{i_1} \psi^{i_2} \dots \psi^{i_{n-2}} \nonumber \\
& & + \phi^a \phi^b A_{a b i_1 i_2 \dots i_{n-4}}(\phi^k) \psi^{i_1} \psi^{i_2} \dots \psi^{i_{n-4}} +  \dots. \
\label{OAfinal}
\end{eqnarray}
Thus,  the globally-defined operator ${\widehat {\cal O}}_A$ is a global section of the sheaf $(\Omega^{ch}_X)^{{\mathfrak t}\geq} \otimes \langle \phi^a \rangle$ of conformal weight $(0,0)$.

Next, we shall proceed to make an important observation about the nature of the $Q$-closed operator ${\widehat {\cal O}}_A$. To this end, let ${\cal O}_A =a, \ {\cal O}^1_A = a_1, \ {\cal O}^2_A = a_2, \dots, {\cal O}^{n /2}_A = a_{n/2}$, where ${\cal O}^k_A$ is the $k^{\textrm{th}}$ correction term added to ${\cal O}_A$ in our final expression of ${\widehat {\cal O}}_A$. Let us denote $ [ (\Omega^{ch}_X)^{\mathfrak t \geq}]^{q-p} \otimes {\langle \phi^a \rangle}^p$ as the subcomplex of $(\Omega^{ch}_X)^{\mathfrak t \geq} \otimes {\langle \phi^a \rangle}$ consisting of elements with $(g_L, g_R)$ ghost number  $(q,p)$. Define $C^{p,q}$ to be any conformal weight $(0,0)$ element of this subcomplex. Then, one can easily see that $a \in C^{0,n}$, $a_1 \in C^{1, n-1}$,  $a_2 \in C^{2, n-2}$ etc. In other words, we can write $a_i \in C^{l + i, h-i}$, where $a_0 = a$, that is, $a \in C^{l,h}$, which then means that $l=0$ and $h=n$. Notice also that if we were to write $\{Q_L , \cal O\}$ and $\{Q_R, \cal O\}$ as ${\tilde d} {\cal O}$ and ${\tilde \delta} {\cal O}$ respectively, from (\ref{spectral1}), (\ref{spectral2}), and the subsequent analogous relations that will follow in our refinement of ${\widehat{\cal O}}_A$, we see that for $a \in C^{l,h}$, we have a system of relations
\begin{eqnarray}
{\tilde d}a & = & 0 \nonumber \\
{\tilde \delta} a & = & - {\tilde d}a_1 \nonumber \\
\label{zigzag}
{\tilde \delta} a_1 & = & - {\tilde d}a_2 \\
{\tilde \delta} a_2 & = & - {\tilde d}a_3 \nonumber \\
& \vdots & \nonumber \
\end{eqnarray}
which admits a solution
\be
(a_1, a_2, \dots) \qquad \textrm{where} \qquad a_i \in C^{l + i, h-i}.
\ee
Thus, (\ref{zigzag}) tells us that an element

\be
{\hat z} : = a\oplus a_1\oplus a_2 \oplus \dots
\ee
lies in $Z^n$, where
\be
{Z^n} : = \{ {\hat z} \in C^n, \ ({\tilde d} + {\tilde \delta}) {\hat z} = 0 \}
\ee
and
$C^n$ is the total double complex defined by
\be
C^n : = \bigoplus_{p+q = n} C^{p,q}
\ee
with a total differential ${\tilde d} + {\tilde \delta} : C^n \to C^{n+1}$, where the individual differentials
\be
 {\tilde d}: C^{p,q} \to C^{p, q+1}, \qquad {\tilde \delta} : C^{p,q} \to C^{p+1, q},
\ee
satisfy
\be
{\tilde d}^2 = 0, \qquad \{{\tilde d},{\tilde \delta} \} = 0, \qquad {\tilde \delta}^2 = 0.
\ee
Since we have $Q^2_L = Q^2_R = 0$, where $Q_L$ and $Q_R$ act to increase $g_L$ and $g_R$ of any physical operator $\cal O$ by one, plus the fact that $\{Q_L, Q_R\} =0$ on $\cal O$, it is clear that one can represent ${\widehat{\cal O}}_A$ by $\hat z$, with $Q_L$ and $Q_R$ corresponding to $\tilde d$ and $\tilde {\delta}$ respectively. Now consider the  system of relations \cite{sternberg}
\begin{eqnarray}
{\tilde d}c_0 + {\tilde \delta}{c_{-1}} & = & b \nonumber \\
{\tilde d}{c_{-1}} + {\tilde \delta}{c_{-2}} & = & 0 \nonumber \\
\label{zigzagboundary}
{\tilde d}{c_{-2}} + {\tilde \delta} {c_{-3}} & = & 0  \\
{\tilde d}{c_{-3}} + {\tilde \delta} {c_{-4}} & = & 0 \nonumber \\
& \vdots & \nonumber \
\end{eqnarray}
where $c_{-i} \in C^{l-i, h+i -1}$, ${\tilde \delta} c_0 = 0$, $b \in B^{l,h} \subset C^{l,h}$, and
\be
B^n : = \bigoplus_{p+q = n} B^{p,q}, \qquad \qquad  B^n : (d+ {\tilde \delta}) C^{n-1}.
\ee
Because $l=0$ and $h=n$, we have $c_{0} \in C^{0, n-1}$, $c_{-1} \in C^{-1, n}$, $c_{-2} \in C^{-2, n+1}$ and so on. Since the local operators cannot have negative $g_R$ values, there are no physical operators corresponding to $c_{-1}$, $c_{-2}$, $c_{-3}$ etc. In other words, there is no solution $(c_0, c_{-1}, c_{-2}, \dots)$ to (\ref{zigzagboundary}), and $B^n$, which consists of the elements $({\tilde d}+ {\tilde \delta}) {\hat b}$, where
\be
{\hat b} : = c_0 \oplus c_{-1} \oplus c_{-2} \oplus \dots \ \in C^{n-1},
\ee
is therefore empty. Consequently, the  cohomology of the double complex $H_{{\tilde d}+{\tilde \delta}} (C^n) = {Z^n / B^n}$, is simply given by $Z^n$: a class in $H_{{\tilde d}+{\tilde \delta}} (C^n)$ can be represented by an element $\hat z$. What this means is that $\widehat{\cal O}_A$, in addition to being $Q$-closed, represents a class in the $Q$-cohomology too, that is, $\widehat{\cal O}_A$ cannot be written as $\{Q, \dots \}$.

Now that we have found our $Q$-closed operator $\widehat{\cal O}_A$, and learnt that it is a class in the $(Q_L +Q_R)$-cohomology, one may then return to our original objective and ask if $\widehat{\cal O}_A$ is part of the subset of operators in the $Q_R$-cohomology which is also in the $Q_L$-cohomology. The answer is yes. This can be explained as follows. Firstly, the system of relations in (\ref{zigzag}) means that the cohomology of the double complex $H_{\tilde d + \tilde \delta} (C^n)$ can be computed using a spectral sequence \cite{sternberg, Bott}. In particular, we have
\be
H_{\tilde d + \tilde \delta} (C^n) = E_{\infty},
\ee
whereby
\begin{eqnarray}
E_1 & = & H_{\tilde d} (C^n), \nonumber \\
E_2 & = & H_{\tilde \delta}H_{\tilde d} (C^n), \nonumber \\
\label{specseq}
E_3 & = & H_{d_2} H_{\tilde \delta}H_{\tilde d} (C^n), \\
& \vdots & \nonumber \\
E_{\infty} & = & H_{d_{\infty}} \dots H_{d_2} H_{\tilde \delta}H_{\tilde d} (C^n). \nonumber \
\end{eqnarray}
More concisely, we have $E_{r+1} = H(E_r, d_r)$, where $E_0 = C^n$, $d_0 = {\tilde d}$, $E_1 = H_{\tilde d}(C^n)$, $d_1 = \tilde \delta$ and so on. Generally, $d_r =0$ for some $r \geq m$, whence the spectral sequence ``collapses at its $E_m$ stage'' and converges to $H_{\tilde d + \tilde \delta} (C^n)$, that is, $E_m = E_{m+1} = \dots =E_{\infty} =H_{\tilde d + \tilde \delta} (C^n)$. Hence, from (\ref{specseq}), we see that any element of $H_{\tilde d + \tilde \delta} (C^n)$ is also an element of $H_{\tilde d} (C^n)$ and $H_{\tilde \delta} (C^n)$. Therefore, $\widehat{\cal O}_A$ represents a class in the $Q_R$- $\it{and}$ $Q_L$-cohomology. In summary, $\widehat{\cal O}_A$ constitutes the subset of conformal weight $(0,0)$ local operators of the half-twisted gauged sigma model which are also in the $Q_L$- and $Q$-cohomology.

How about the higher conformal weight operators? Let us begin with a general weight (1,0) operator
\be
{\cal O}_B = B^j_{i_1 i_2 \dots i_n}(\phi^k) p_{zj} \psi^{i_1} \psi^{i_2} \dots \psi^{i_n}
\label{OB}
\ee
which may be admissible  in the $Q_R$-cohomology. (As before, we have not included the $\phi^a$ field in ${\cal O}_B$ because it will soon appear in our discussion.)  Let us denote $\Delta {\cal O}_B$ as the change in ${\cal O}_B$ due to the action of $Q$, that is,
\be
\Delta {\cal O}_B = \{Q_L, {\cal O}_B \} + \{ Q_R, {\cal O}_B \}.
\label{deltaOB}
\ee
As in our discussion on ${\cal O}_A$, let us choose ${\cal O}_B$ such that it can be annihilated by $Q_L$, that is, $\{Q_L, {\cal O}_B\} = 0$, so that it may be admissible as a class in the $Q_L$-cohomology as well. Then,
\begin{eqnarray}
\Delta {\cal O}_B & = & \{ Q_R, {\cal O}_B \} \nonumber \\
&=& -i n \phi^a V_a^{i_1} B^j_{i_1 i_2 \dots i_n} p_{zj} \psi^{i_2} \dots \psi^{i_n}. \
\end{eqnarray}
Thus,     as in the     case with ${\cal O}_A$, we find that ${\cal O}_B$ is neither in the $Q_R$-cohomology nor $Q$-closed as required. These observations suggest that corrections to the operator ${\cal O}_B$ need to be made. To this end, recall from our discussion above on ${\cal O}_a$, that if ${\cal O}_B$ is to    be   admissible as an operator and is $Q_L$-closed, we may have
\begin{eqnarray}
\{Q_R, {\cal O}_B\} & = & - \{Q_L, {\cal O}^1_B\}\nonumber \\
& = & -i n \phi^a V_a^{i_1}  B^j_{i_1 i_2 \dots i_n} p_{zj} \psi^{i_2} \dots \psi^{i_n}, \
\label{spectralb1}
\end{eqnarray}
so that one may `refine' the definition of ${\cal O}_B$ to
\begin{eqnarray}
{\widehat {\cal O}}_B  & =  & {\cal O}_B + {\cal O}^1_B \nonumber \\
&=& B^j_{i_1 i_2 \dots i_n} (\phi^k) p_{zj} \psi^{i_1} \psi^{i_2} \dots \psi^{i_n} + \phi^a B^j_{a i_1 i_2 \dots i_{n-2}}(\phi^k) p_{zj} \psi^{i_1} \psi^{i_2} \dots \psi^{i_{n-2}}, \
\label{OB1}
\end{eqnarray}
where
\be
\partial_m B^j_{a i_1 i_2 \dots i_{n-2}} p_{zj}  \psi^m \psi^{i_1}\psi^{i_2} \dots \psi^{i_{n-2}} = n V_a^{i_1} B^j_{i_1 i_2 i_3 \dots i_n} p_{zj} \psi^{i_2} \psi^{i_3} \dots \psi^{i_n},
\label{b1}
\ee
and so on,  just as we did to derive the final form of ${\widehat {\cal O}}_A$. However, since $p_i$, or alternatively $\beta_i$, transforms in a complicated fashion over an intersection of open sets $U_1 \cap U_2$ in $X$ \cite{MSV1,MC}, ${\cal O}_B$ may not be globally well-defined. Likewise for ${\cal O}^1_B$. Hence, these operators are not admissible as global sections of the sheaves $(\Omega^{ch}_X)^{\mathfrak t \geq}$ or $(\Omega^{ch}_X)^{\mathfrak t \geq}\otimes \langle \phi^a\rangle$ in general. Thus, in contrast to ${\widehat{\cal O}}_A$, we do not have a consistent procedure to define ${\widehat{\cal O}}_B$ as a class in $H_{\tilde d + \tilde \delta}(C^n)$. In other words, operators which are admissible in the $Q$- and hence $Q_R$- and $Q_L$-cohomology, cannot contain the $p_i$ fields or their higher $z$-derivatives.

Another weight $(1,0)$ operator   that one can consider is
\be
{\cal O}_C = C^k_{i_1 i_2 \dots i_n}(\phi^j) \psi_{zk} \psi^{i_1} \psi^{i_2} \dots \psi^{i_n}
\label{OC}
\ee
which may be admissible  in the $Q_R$-cohomology. (Again, we have not included the $\phi^a$ field in ${\cal O}_C$ because it will appear in our following discussion.)  Let us denote $\Delta {\cal O}_C$ as the change in ${\cal O}_C$ due to the action of $Q$, that is,
\be
\Delta {\cal O}_C = \{Q_L, {\cal O}_C \} + \{ Q_R, {\cal O}_C \}.
\label{deltaOC}
\ee
As in our previous examples, let us choose ${\cal O}_C$ such that it can be annihilated by $Q_L$, that is, $\{Q_L, {\cal O}_C\} = 0$, so that it may be admissible as a class in the $Q_L$-cohomology as well. Then,
\begin{eqnarray}
\Delta {\cal O}_C & = & \{ Q_R, {\cal O}_C \} \nonumber \\
&=& -i n \phi^a V_a^{i_1} C^k_{i_1 i_2 \dots i_n} \psi_{zk} \psi^{i_2} \dots \psi^{i_n}. \
\end{eqnarray}
Unlike $p_i$, the field $\psi_{zk}$ does not have a complicated transformation law over an intersection of open sets $U_1 \cap U_2$ in $X$ \cite{MSV1,MC}. Thus, ${\cal O}_C$ can correspond to a global section of $(\Omega^{ch}_X)^{\mathfrak t \geq}$. Recall from our discussion on ${\cal O}_a$ that we can write
\begin{eqnarray}
\{Q_R, {\cal O}_C\} & = & - \{Q_L, {\cal O}^1_C\}\nonumber \\
& = & -i n \phi^a V_a^{i_1}  C^k_{i_1 i_2 \dots i_n} \psi_{zk} \psi^{i_2} \dots \psi^{i_n}, \
\label{spectralc1}
\end{eqnarray}
so that one may `refine' the definition of ${\cal O}_C$ to
\begin{eqnarray}
{\widehat {\cal O}}_C  & =  & {\cal O}_C + {\cal O}^1_C, \
\label{OC1}
\end{eqnarray}
just as we did for ${\widehat{\cal O}}_A$ and ${\widehat{\cal O}}_B$, and so on. However, from (\ref{deltaL1})-(\ref{deltaL2}), we have $\delta_L \psi_{zi} = -p_{zi}$ and $\delta_L p_{zi} = 0$, and a little thought reveals that there are no weight $(1,0)$ operators ${\cal O}^1_C$ which can satisfy (\ref{spectralc1}). Thus, the construction fails and one cannot proceed to make further corrections to ${\cal O}_C$. In other words, operators which are admissible in the $Q$- and hence $Q_R$- and $Q_L$-cohomology, cannot contain the $\psi_{zk}$ fields or their higher $z$-derivatives.

In fact, the above observations about higher weight operators in the last two paragraphs, are consistent with the results of sect. 3.4 which states that because $T_{zz}$ is $Q_L$-exact, that is, $T_{zz} = \{Q_L, G_{zz}\}$ for some operator $G_{zz}$, an operator in the $Q_L$-cohomology must be of weight $(0, m)$ for $m\geq 0$. Since $p_{zi}$, $\psi_{zk}$ and their higher $z$-derivatives are of weight $(l, 0)$ where $l \geq 1$, they cannot be included in an operator that is admissible. Likewise, we cannot have the field $\psi^a_z$, its higher $z$-derivatives, and the higher $z$-derivatives of the fields $\phi^i$, $\phi^a$ and $\psi^i$ either.

\vspace{0.4cm}{\noindent \it {The Chiral Equivariant Cohomology ${\bf{H}}_{T^d}(\Omega^{ch}_X)$}}

In rewriting $Q_L(z)$ (as given in (\ref{supercurrents})) in terms of the $\beta_i(z)$ and $c^i(z)$ fields, we see that $Q_L$ coincides with $d_{\cal Q}$, the differential of the chiral de Rham complex $\Omega^{ch}_X$ on $X$ \cite{MSV1}.\footnote{The differential $d_{\cal Q}$ in \cite{MSV1} is actually $-Q_L$ because of a trivial sign difference in defining $\beta_i(z)$. However, the sign convention adopted for $\beta_i(z)$ in this paper is the same as in \cite{andy1}, which is our main point of interest.} Another observation to be made is that $Q_R(z)$ (as given in (\ref{supercurrents})) can be written as $-\phi^a \iota_{V_a}(z)$, where $\iota_{V_a}(z) = V_a^i \psi_{zi}(z)$ is just a vertex algebraic analogue of the interior product by the holomorphic vector field $V_a$ on $X$. Indeed, after rewriting $\iota_{V_a}(z)$ in terms of the $\gamma^i(z)$ and $b_i(z)$ fields, one can compute its OPE with $L_{V_a}(z)$ (given in (\ref{LV})) as
\be
L_{V_a}(z) \iota_{V_b}(z') \sim {\iota_{[V_a, V_b]}(z') \over {z-z'}}.
\label{ope1}
\ee
Moreover, one can also compute that
\be
\iota_{V_a} (z) \iota_{V_b} (z') \sim 0.
\label{ope2}
\ee
Clearly, (\ref{ope1}) and (\ref{ope2}) are just the vertex algebraic analogue of the differential-geometric relations $[L_{\xi}, \iota_{\eta}] = \iota_{[\xi, \eta]}$ and $\{\iota_{\xi}, \iota_{\eta}\} =0$ respectively, where $\xi$ and $\eta$ are any two vector fields on $X$. Since $\iota_{V_a}(z)$ can only consist of the $\phi^i$($\gamma^i$) and $\psi_{zi}$($b_i$) fields in general, it must be a section of the sheaf $\Omega^{ch}_X$. Now recall that $\psi_{zi}$ transforms as a section of $\Phi^*(T^*X)$  on $\Sigma$, that is, over an arbitrary intersection $U_1 \cap U_2$ in $X$, we have the transformation ${\tilde \psi}_{zj} (z) = {\psi}_{zi} {\partial \phi^i \over {\partial {\tilde \phi}^j}}(z)$. On the other hand, a holomorphic vector such as $V_a^i(z)$ will transform as ${\tilde V}_a^j(z) ={V}_a^i {\partial {\tilde \phi}^j \over {\partial {\phi}^i}}(z)$. This means that over an arbitrary intersection $U_1 \cap U_2$ in $X$, we have ${\tilde V}_a^i (z) {\tilde \psi}_{zi}(z) =  V_a^j(z) \psi_{zj}(z)$. This can be written in terms of the $\gamma^i(z)$ and $b_i(z)$ fields as
\be
{\tilde V}_a^i (z) {\tilde b}_{i}(z) =  V_a^j(z) b_{j}(z),
\ee
that is, ${\tilde\iota}_{V_a} (z) = \iota_{V_a}(z)$. This means that the conformal weight $(1,0)$ vertex operator $\iota_{V_a}(z)$ must be a global section of the sheaf $\Omega^{ch}_X$.

Finally, notice that the sheaf $(\Omega^{ch}_X)^{\mathfrak t \geq} \otimes {\langle \phi^a \rangle}$ coincides with the $\it{small}$ chiral Cartan complex ${\cal C}_{T^d}(\Omega^{ch}_X)$ defined by Lian et al. in sect. 6.2 of \cite{andy1}. Moreover, via the discussion above and (\ref{supercharges}), we see that the BRST operator $Q= Q_L + Q_R$ can be written as $d_{T^d}= d_{\cal Q} - (\phi^a \iota_{V_a}) (0)$, where  $(\phi^a \iota_{V_a} ) (0) = \oint {dz \over {2\pi i}} \phi^a \iota_{V_a}(z)$. Hence, $d_{T^d}$ coincides with the differential of ${\cal C}_T(\Omega^{ch}_X)$ defined in sect. 6.2 of \cite{andy1}. Therefore, ${\widehat {\cal O}}_A$ represents a class in $H({\cal C}_{T^d}(\Omega^{ch}_X), d_{T^d})$, the $d_{T^d}$-cohomology of the small chiral Cartan complex. From Theorem 6.5 of \cite{andy1}, we have, for any $T^d$-manifold, the isomorphism ${\bf{H}}_{T^d} (\Omega^{ch}_X) \cong H({\cal C}_{T^d}(\Omega^{ch}_X), d_{T^d})$, where ${\bf{H}}_{T^d} (\Omega^{ch}_X)$ is the $T^d$-equivariant cohomology of the chiral de Rham complex. Thus, ${\widehat{\cal O}}_A$ actually represents a conformal weight (0,0) class in ${\bf{H}}_{T^d} (\Omega^{ch}_X)$! In addition, from the discussion in the last few paragraphs on the vanishing of other operators in the  $Q$-cohomology, we learn that the only classes in ${\bf{H}}_{T^d} (\Omega^{ch}_X)$ are represented by the operators ${\widehat{\cal O}}_A$.  Hence, for $G=T^d$, the chiral equivariant cohomology can be described by the subset of physical operators of the half-twisted gauged sigma model which also belong in the $Q_L$-cohomology. In fact, via this description of the chiral equivariant cohomology in terms of a two-dimensional sigma model, the mathematical result in Corollary 6.4 of \cite{andy2} stating that there are $\it{no}$ positive weight classes in ${\bf{H}}_{T^d} (\Omega^{ch}_X)$, now lends itself to a simple and purely physical explanation. In particular, since the holomorphic stress tensor is $Q_L$-exact, that is, $T_{zz} = \{Q_L, G_{zz}\}$ for some operator $G_{zz}$, the physical operators in the $Q_L$-cohomology must be of conformal weight $(0,m)$ for $m \geq 0$. On the other hand, since the antiholomorphic stress tensor is $Q_R$-exact, that is, $T_{\bar z \bar z} = \{Q_R, G_{\bar z \bar z}\}$ for some operator $G_{\bar z \bar z}$, the physical operators in the $Q_R$-cohomology must be of conformal weight $(n,0)$ for $n \geq 0$. Therefore, the physical operators in the $Q$-cohomology, which we have shown earlier to correspond to operators that are also in the $Q_L$- and $Q_R$-cohomology, must be of conformal weight $(0,0)$, that is, they must be ground operators. Since these operators of the $Q$-cohomology represent the only classes in ${\bf{H}}_{T^d} (\Omega^{ch}_X)$, there are consequently no classes of positive weight in ${\bf{H}}_{T^d} (\Omega^{ch}_X)$.

Last but not least, that $H_{\tilde d + \tilde\delta }(C^n)$ and therefore $H({\cal C}_{T^d}(\Omega^{ch}_X), d_{T^d})$ can be constructed via a converging spectral sequence $(E_r, d_r)$ which collapses at $E_r$ for some $r$, is also consistent with Theorem 6.6 of \cite{andy1}. Thus, the chiral equivariant cohomology can indeed be consistently represented by the ground operators of a two-dimensional half-twisted gauged sigma model.

\newsubsection{Correlation Functions and Topological Invariants}

In this subsection, we shall examine the correlation functions of local operators of type ${\widehat{\cal O}}_A$. We will also define some non-local operators in the $Q$-cohomology and study their correlation functions as well. In doing so, we shall be able to derive a set of topological invariants on $X$. These invariants can then be used to provide a purely physical verification of the isomorphism between the weight-zero subspace of ${\bf{H}}_{T^d}( \Omega^{ch}_X)$ and the classical equivariant cohomology of $X$ \cite{andy1, andy2}.

\vspace{0.4cm}{\noindent \it {Local Operators}}

To begin with, let $P_1, P_2, \dots, P_k$ be $k$ distinct points on $\Sigma$. Let ${\cal O}_{A_1}, {\cal O}_{A_2}, \dots, {\cal O}_{A_K}$ be local operators of type ${\cal O}_A$ with $n_1, n_2, \dots, n_k$ number of $\psi^i$ fields. Let ${\widehat {\cal O}}_{A_1}, {\widehat {\cal O}}_{A_2}, \dots, {\widehat {\cal O}}_{A_K}$ be the corresponding operators which represent classes in ${\bf H}_{T^d} (\Omega^{ch}_X)$. Consider a non-vanishing correlation function of such operators (where $\Sigma$ is a simply-connected, genus-zero Riemann surface  in perturbation theory):
\be
Z(A_1, A_2, \dots, A_K) = {\langle {\widehat {\cal O}}_{A_1}(P_1) {\widehat {\cal O}}_{A_2}(P_2) \dots {\widehat {\cal O}}_{A_K} (P_K) \rangle}_0.
\label{correlation local}
\ee
$Z(A_1, A_2, \dots, A_K)$ is a topological invariant in the sense that it is invariant under changes in the metric and complex structure of $\Sigma$ or $X$. Indeed, since ${\cal L}_{\textrm{gauged}} = \{Q, V_{\textrm{gauged}} \}$, a change in the metric and complex structure of $\Sigma$ or $X$ will result in a change in the Lagrangian $\delta {\cal L} = \{Q, V'\}$ for some $V'$. Hence, because $\{Q, {\widehat {\cal O}}_{A_i}(P_i)\} = 0$, and $\langle \{Q, Y\} \rangle = 0$ for any operator $Y$, the corresponding change in $Z(A_1, A_2, \dots, A_K)$ will be given by
\begin{eqnarray}
\delta Z &  = & {\langle {\widehat {\cal O}}_{A_1} {\widehat {\cal O}}_{A_2}\dots {\widehat {\cal O}}_{A_K} (-\delta {\cal L}) \rangle}_0 \nonumber \\
& =& -  {\langle {\widehat {\cal O}}_{A_1}{\widehat {\cal O}}_{A_2} \dots {\widehat {\cal O}}_{A_K}  \{Q, V'\} \rangle}_0 \nonumber \\
& =& - {\langle \{Q,  \Pi_i   {\widehat {\cal O}}_{A_i} \cdot V'  \} \rangle}_0 \nonumber \\
& = & 0. \
\end{eqnarray}

\vspace{1.5cm}{\noindent \it {Non-Local Operators}}

We shall now continue to construct the non-local operators of the theory, that is, operators which are globally-defined on $\Sigma$. Unlike ${\widehat {\cal O}}_{A_i}$ above, these operators will not define a chiral algebra $\cal A$. (Recall from the discussion at the end of sect. 3.5, that a chiral algebra must be locally-defined on $\Sigma$ unless $\Sigma$ is of genus one).  However, they will correspond to classes in ${\bf H}_{T^d} (\Omega^{ch}_X)$, as we will see.

To this end, notice that we can always view ${\widehat {\cal O}}_A$ as an operator-valued zero-form on $\Sigma$. Let us then rewrite it as ${\widehat {\cal O}^{(0)}}_A$, where the superscript $(0)$ just denotes that the operator is a zero-form on $\Sigma$. Let us now try to compute the exterior derivative of ${\widehat {\cal O}^{(0)}}_A$ on $\Sigma$
\be
d{\widehat {\cal O}^{(0)}}_A = \partial_z {\widehat {\cal O}^{(0)}}_A dz + \partial_{\bar z} {\widehat {\cal O}^{(0)}}_A d{\bar z}.
\label{first exterior derivative}
\ee
(The motivation for doing so will be clear shortly). The partial $z$-derivative will be given by
\begin{eqnarray}
\partial_z {\widehat {\cal O}^{(0)}}_A & = & {\partial \over {\partial z}} ( A_{i_1 i_2 \dots i_n} \psi^{i_1} \psi^{i_2} \dots \psi^{i_n} + \phi^a A_{a i_1 i_2 \dots i_{n-2}} \psi^{i_1} \psi^{i_2} \dots \psi^{i_{n-2}} \nonumber \\
&  & \qquad  + \phi^a \phi^b A_{a b i_1 i_2 \dots i_{n-4}}\psi^{i_1} \psi^{i_2} \dots \psi^{i_{n-4}} + \dots ) \nonumber \\
 \nonumber \\
& = & \partial_k A_{i_1 i_2 \dots i_n}\partial_z \phi^k \psi^{i_1} \psi^{i_2} \dots \psi^{i_n} + n A_{i_1 i_2 \dots i_n} \partial_z \psi^{i_1} \psi^{i_2} \dots \psi^{i_n}  \nonumber \\
&& + \partial_z \phi^a A_{a i_1 i_2 \dots i_{n-2}} \psi^{i_1} \psi^{i_2} \dots \psi^{i_{n-2}} + \phi^a \partial_k A_{a i_1 i_2 \dots i_{n-2}} \partial_z \phi^k \psi^{i_1} \psi^{i_2} \dots \psi^{i_{n-2}} \nonumber \\
&& + (n-2) \phi^a A_{a i_1 i_2 \dots i_{n-2}} \partial_z \psi^{i_1} \psi^{i_2} \dots \psi^{i_{n-2}} + \dots \nonumber \\
\label{partial1}
 \\
& = & n A_{i_1 i_2 \dots i_n} \partial_z \psi^{i_1} \psi^{i_2} \dots \psi^{i_n}  + \partial_z \phi^a A_{a i_1 i_2 \dots i_{n-2}} \psi^{i_1} \psi^{i_2} \dots \psi^{i_{n-2}} \nonumber \\
&& + \phi^a \partial_k A_{a i_1 i_2 \dots i_{n-2}} \partial_z \phi^k \psi^{i_1} \psi^{i_2} \dots \psi^{i_{n-2}}+ (n-2) \phi^a A_{a i_1 i_2 \dots i_{n-2}} \partial_z \psi^{i_1} \psi^{i_2} \dots \psi^{i_{n-2}}  \nonumber \\
&& + \dots, \nonumber \
\end{eqnarray}
where the condition $\{Q_L, {\cal O}_A \} = 0$ implies that for our purpose, one can discard the first term on the right-hand side of the second equality in (\ref{partial1}) to arrive at the final equality, since it will not contribute to $d{\widehat {\cal O}^{(0)}}_A$ in (\ref{first exterior derivative}).\footnote{From the field variations $\delta_L \phi^i = \psi^i$ and $\delta_L \psi^i = 0$, the expression $Q_L(z) = \oint{dz \over {2 \pi i}}p_{zi}\psi^i (z)$, and the operator product expansion $p_{zi}(z) \phi^i(z') \sim (z-z')^{-1}$, one can see that $Q_L$ acts on ${\cal O}_A$ as the exterior derivative $d\phi^k {\partial \over{\partial \phi^k}}$. Noting that $d \phi^k = \partial_z \phi^k dz + \partial_{\bar z} \phi^k d{\bar z} = \partial_z \phi^k dz$ since $\partial_{\bar z} \phi^k = 0$, one will have $\{Q_L, {\cal O}_A\}= \partial_k A_{i_1 i_2 \dots i_n} d\phi^k \psi^{i_1} \psi^{i_2} \dots \psi^{i_n} =  \partial_k A_{i_1 i_2 \dots i_n} \partial_z \phi^k dz\ \psi^{i_1} \psi^{i_2} \dots \psi^{i_n}  = 0$. This then implies that one can discard the term $\partial_k A_{i_1 i_2 \dots i_n} \partial_z \phi^k \psi^{i_1} \psi^{i_2} \dots \psi^{i_n}$ in computing $\partial_z {\widehat {\cal O}^{(0)}}_A $, since it vanishes in $\partial_z {\widehat {\cal O}^{(0)}}_A dz$.} The $\bar z$-derivative will be given by
\begin{eqnarray}
\partial_{\bar z} {\widehat {\cal O}^{(0)}}_A & = & {\partial \over {\partial {\bar z}}} ( A_{i_1 i_2 \dots i_n} \psi^{i_1} \psi^{i_2} \dots \psi^{i_n} + \phi^a A_{a i_1 i_2 \dots i_{n-2}} \psi^{i_1} \psi^{i_2} \dots \psi^{i_{n-2}} \nonumber \\
&& \quad \quad + \phi^a \phi^b A_{a b i_1 i_2 \dots i_{n-4}}\psi^{i_1} \psi^{i_2} \dots \psi^{i_{n-4}} + \dots ) \nonumber \\
\nonumber \\
& = & \partial_k A_{i_1 i_2 \dots i_n}\partial_{\bar z} \phi^k \psi^{i_1} \psi^{i_2} \dots \psi^{i_n} + n A_{i_1 i_2 \dots i_n} \partial_{\bar z} \psi^{i_1} \psi^{i_2} \dots \psi^{i_n}  \nonumber \\
&& + \partial_{\bar z} \phi^a A_{a i_1 i_2 \dots i_{n-2}} \psi^{i_1} \psi^{i_2} \dots \psi^{i_{n-2}} + \phi^a \partial_k A_{a i_1 i_2 \dots i_{n-2}} \partial_{\bar z} \phi^k \psi^{i_1} \psi^{i_2} \dots \psi^{i_{n-2}} \nonumber \\
&& + (n-2) \phi^a A_{a i_1 i_2 \dots i_{n-2}} \partial_{\bar z} \psi^{i_1} \psi^{i_2} \dots \psi^{i_{n-2}} + \dots \nonumber \\
\label{partial2}
\\
& = & 0, \nonumber \
\end{eqnarray}
where we have used the equations of motion $\partial_{\bar z} \phi^k = \partial_{\bar z} \psi^i = 0$ and the fact that $\partial_{\bar z} \phi^a = 0$,\footnote{Note from discussion in sect. 3.5 that any operator $\cal O$ in the $Q_R$-cohomology varies holomorphically with $z$. Since, ${\widehat{\cal O}}_A$ is such an operator, and it contains the fields $\phi^i$, $\psi^i$ and $\phi^a$, where $\phi^i$ and $\psi^i$ are holomorphic in $z$ from the equations of motion, we deduce that $\phi^a$ must be holomorphic in $z$ as well.} in going from the second to third equality in (\ref{partial2}). Hence, we can write
\begin{eqnarray}
d{\widehat {\cal O}^{(0)}}_A & = &  \partial_z {\widehat {\cal O}^{(0)}}_A dz \nonumber \\
&= & n A_{i_1 i_2 \dots i_n} d\psi^{i_1} \psi^{i_2} \dots \psi^{i_n}  + d\phi^a A_{a i_1 i_2 \dots i_{n-2}} \psi^{i_1} \psi^{i_2} \dots \psi^{i_{n-2}} \nonumber \\
&& + \phi^a \partial_k A_{a i_1 i_2 \dots i_{n-2}} d\phi^k \psi^{i_1} \psi^{i_2} \dots \psi^{i_{n-2}}+ (n-2) \phi^a A_{a i_1 i_2 \dots i_{n-2}} d \psi^{i_1} \psi^{i_2} \dots \psi^{i_{n-2}}  \nonumber \\
\label{doa1}
&& + \dots  \
\end{eqnarray}
In fact, one can show that
\be
d{\widehat {\cal O}^{(0)}}_A  = \{ Q, {\widehat {\cal O}^{(1)}}_A \},
\label{doa2}
\ee
whereby ${\widehat {\cal O}^{(1)}}_A$ is an operator-valued one-form on $\Sigma$. For ease of illustration, let us take ${\widehat {\cal O}^{(0)}}_A$ to be of type $n=2$, that is,
\be
{\widehat {\cal O}^{(0)}}_A = A_{i_1 i_2} \psi^{i_1} \psi^{i_2}  + \phi^a A_{a}.
\label{eg}
\ee
Then, from (\ref{doa1}), we find that
\be
d{\widehat {\cal O}^{(0)}}_A = 2 A_{i_1 i_2} d\psi^{i_1} \psi^{i_2}  + d\phi^a A_{a} + \phi^a \partial_k A_{a} d\phi^k.
\ee
But from (\ref{c1}), and the identification of $\psi^i$ as $d\phi^i$ as explained in footnote 17, we have the condition
\be
\partial_k A_a d\phi^k = 2 V_a^{i_1} A_{i_1 i_2} d\phi^{i_2},
\label{condition}
\ee
so that
\be
d{\widehat {\cal O}^{(0)}}_A = 2 A_{i_1 i_2} \psi^{i_1} d\psi^{i_2}  + 2 \phi^a V_a^{i_1}A_{i_1 i_2} d\phi^{i_2}  + d\phi^a A_{a}.
\ee
Next, from  (\ref{condition}), we deduce that $\partial_k A_a = 2 V_a^{i_1} A_{i_1 k}$ for $k = 1,2, \dots, \textrm{dim}_{\mathbb C}X$. In order to satisfy the condition $\{Q_L, {\cal O}_A\} = 0$, one can simply choose $\partial_l A_{i_1 i_2} = 0$ or $A_{i_1 i_2}$ constant. (The present discussion can be generalised to non-constant $A_{i_1 i_2}$ as will be explained shortly). And since $\partial_{\bar l} V_a^i = \partial_l V_a^i = 0$ for abelian $G=T^n$, we can thus write $A_a$ as
\be
A_a = 2 \sum^{\textrm{dim}_{\mathbb C}X}_{\alpha =1}  V_a^{i_1} A_{i_1 \alpha} \phi^{\alpha}.
\ee
If we let
\be
{\hat A}_a = 2 \sum^{\textrm{dim}_{\mathbb C}X}_{j=1}  ({\phi^b}V_b^j)^{-1} V_a^{i_1} A_{i_1 j} \phi^j \psi^j,
\ee
one can verify that we will indeed have $d{\widehat {\cal O}^{(0)}}_A  = \{ Q, {\widehat {\cal O}^{(1)}}_A \}$, where
\be
{\widehat {\cal O}^{(1)}}_A = 2 i A_{i_1 i_2} \psi^{i_1} d\phi^{i_2}  + i d \phi^a {\hat A}_{a}.
\ee
One can use similar arguments to show that (\ref{doa2}) holds for ${\widehat {\cal O}^{(0)}}_A$ of type $n > 2$ as well. Consequently, one can go further to define the non-local operator
\be
W_A (\zeta) = \int_{\zeta} {\widehat {\cal O}^{(1)}}_A,
\ee
such that if $\zeta$ is a homology one-cycle on $\Sigma$, (i.e. $\partial \zeta = 0$), then
\be
\{Q, W_A(\zeta)\}  =  \int_{\zeta} \{Q, {\widehat {\cal O}^{(1)}}_A \}  =   \int_{\zeta} d{\widehat {\cal O}^{(0)}}_A = 0, \\
\ee
that is, $W_A(\zeta)$ is a $Q$-invariant operator.

One can also deduce the relation $d{\widehat {\cal O}^{(0)}}_A  = \{ Q, {\widehat {\cal O}^{(1)}}_A \}$ via the following argument. Firstly, note that since $Z(A_1, A_2, \dots, A_K) = {\langle {\widehat {\cal O}}^{(0)}_{A_1}(P_1) {\widehat {\cal O}}^{(0)}_{A_2}(P_2) \dots {\widehat {\cal O}}^{(0)}_{A_K} (P_K) \rangle}_0$ is a topological invariant in that it is independent of changes in the metric and complex structure of $\Sigma$ or $X$, it will mean that it is invariant under changes in the points of insertion $P_1, P_2, \dots, P_k$, that is,
\be
{\left \langle \left ({\widehat {\cal O}}^{(0)}_{A_1}(P'_1) - {\widehat {\cal O}}^{(0)}_{A_1}(P_1) \right) {\widehat {\cal O}}^{(0)}_{A_2}(P_2) \dots {\widehat {\cal O}}^{(0)}_{A_K} (P_K) \right \rangle}_0 = 0,
\ee
or rather
\be
{\left \langle \left (\int_{\zeta} d {\widehat {\cal O}}^{(0)}_{A_1}\right){\widehat {\cal O}}^{(0)}_{A_2}(P_2) \dots {\widehat {\cal O}}^{(0)}_{A_K} (P_K) \right \rangle}_0 = 0,
\ee
where $\zeta$ is a path that connects $P'_1$ to $P_1$ on $\Sigma$. Since $\{Q, Y\} = 0$ for any operator $Y$, and since $\{Q, {\widehat {\cal O}}^{(0)}_{A_i} \} =0$ for any $i=1, 2, \dots, k$, it must be true that
\be
\int_{\zeta} d {\widehat {\cal O}}^{(0)}_{A_1} = \{Q, W_{A_1}(\zeta) \},
\label{ss}
\ee
and for consistency with the left-hand side of (\ref{ss}), $W_{A_1}(\zeta)$ must be an operator-valued zero-form on $\Sigma$ that depends on $\zeta$, and where its explicit form will depend on ${\cal O}_{A_1}$. Such a non-local operator can be written as $W_{A_1}(\zeta) = \int_{\zeta}{\widehat {\cal O}}^{(1)}_{A_1}$, where ${\widehat {\cal O}}^{(1)}_{A_1}$ is an operator-valued one-form on $\Sigma$, and its explicit form depends on ${\cal O}_{A_1}$. Hence, from (\ref{ss}), it will mean that
\be
d{\widehat {\cal O}^{(0)}}_A  = \{ Q, {\widehat {\cal O}^{(1)}}_A \}
\ee
as we have illustrated with an example earlier. (Note that because the above arguments hold in all generality, one can replace ${\widehat {\cal O}}^{(0)}_{A}$ in (\ref{eg}) with another consisting of a non-constant $A_{i_1 i_2}$, and still illustrate that the relation in (\ref{doa2}) holds).

Let us now consider the correlation function of $k$ $Q$-invariant operators $W_A(\zeta)$:
\be
Z\left ((A_1, \zeta_1), (A_2, \zeta_2), \dots, (A_k, \zeta_k) \right ) ={ \langle W_{A_1}(\zeta_1) \dots W_{A_k} (\zeta_k) \rangle}_0.
\ee
Under a variation in the metric of $\Sigma$ or $X$, we have
\begin{eqnarray}
\delta Z  & = & {\langle W_{A_1}(\zeta_1) \dots W_{A_k} (\zeta_k) (-\delta{\cal L}) \rangle}_0 \nonumber \\
& = & {\langle W_{A_1}(\zeta_1) \dots W_{A_k} (\zeta_k) \{Q, V'\} \rangle}_0 \nonumber \\
& = &  {\langle \{ Q, \Pi _{i=1}^{k} W_{A_i}(\zeta_i) \cdot V' \} \rangle}_0, \nonumber \\
& = & 0,\
\end{eqnarray}
where we have used $\{Q, W_{A_i} (\zeta_i)\} = 0$, and $\{Q, Y\} = 0$ for any operator $Y$. This means that $Z\left ((A_1, \zeta_1), (A_2, \zeta_2), \dots, (A_k, \zeta_k) \right)$ is a topological invariant, and is independent of changes in the metric and complex structure of $\Sigma$ and $X$. Hence, it will be true that
\be
{\left \langle \ \left [W_{A_1}(\zeta_1) - W_{A_1}(\zeta'_1) \right] W_{A_2} (\zeta_2) \dots W_{A_k} (\zeta_k) \ \right \rangle}_0 = 0,
\label{zz}
\ee
where $\zeta'_1$ is a small displacement of $\zeta_1$, and both are homology one-cycles on $\Sigma$. Define $\zeta_1$ and $\zeta'_1$  to have opposite orientations such that they link a two-dimensional manifold $S$ in $\Sigma$. Then, we will have
\be
W_{A_1}(\zeta_1) - W_{A_1}(\zeta'_1)  = \int_{\zeta_1} {\widehat {\cal O}}^{(1)}_{A_1}(\zeta_1) - \int_{\zeta'_1} {\widehat {\cal O}}^{(1)}_{A_1}(\zeta'_1) = \int_S d {\widehat {\cal O}}^{(1)}_{A_1},
\ee
and from (\ref{zz}), we deduce that
\be
\int_S  d{\widehat {\cal O}}^{(1)}_{A_1} =  \{Q, W_{A_1}(S) \},
\label{sss}
\ee
where again, to be consistent with the left-hand side of (\ref{sss}), $W_{A_1}(S)$ must be an operator-valued zero-form on $\Sigma$, where its explicit form will depend on ${\cal O}_{A_1}$ and $S$. Such a non-local operator can be written as
\be
W_{A_1} (S) = \int_S {\widehat {\cal O}}^{(2)}_{A_1},
\ee
where ${\widehat {\cal O}}^{(2)}_{A_1}$ is an operator-valued two-form on $\Sigma$, and its explicit form depends on ${\cal O}_{A_1}$. Thus, we can write
\be
d{\widehat {\cal O}}^{(1)}_A  = \{Q, {\widehat {\cal O}}^{(2)}_A \}.
\ee
This implies that $W_A(\zeta) = \int_{\zeta}{\widehat {\cal O}}^{(1)}_A$ depends only on the homology class that $\zeta$ represents. Indeed, if $\zeta = \partial \eta$ for some two-manifold $\eta$ in $\Sigma$, we will have
\be
W_A(\zeta) = \int_{\zeta} {\widehat {\cal O}}^{(1)}_A = \int_{\eta} d{\widehat {\cal O}}^{(1)}_A = \{Q, \int_{\zeta} {\widehat {\cal O}}^{(2)}_A\},
\ee
that is, $W_A(\zeta)$ vanishes in $Q$-cohomology if $\zeta$ is trivial in homology. And since $\Sigma$ has real complex dimension 2, it cannot support forms of degree higher than two. Hence,
\be
d{\widehat {\cal O}}^{(2)}_A = 0.
\ee
Let us now define the non-local operator
\be
W_A(\Sigma) = \int_{\Sigma} {\widehat {\cal O}}^{(2)}_A,
\ee
where $\Sigma$ is the worldsheet Riemann surface which is therefore a homology two-cycle because $\partial \Sigma = 0$. Consequently, we have
\be
\{Q, W_A(\Sigma) \} = \int_{\Sigma} \{Q, {\widehat {\cal O}}^{(2)}_A \} = \int_{\Sigma} d{\widehat {\cal O}}^{(1)}_A = \int_{\partial \Sigma} {\widehat {\cal O}}^{(1)}_A = 0,
\ee
that is, $W_A(\Sigma)$ is $Q$-invariant. Hence, correlation functions involving the operators $W_A(P)$, $W_A(\zeta)$ and $W_A(\Sigma)$, will also be invariant under a variation in the metric of $\Sigma$ or $X$.

In summary, we have the local operator
\be
W_A(P) = {\widehat {\cal O}}^{(0)}_A,
\ee
where $P$ is just a zero-cycle or a point on $\Sigma$, and the non-local operators
\be
W_A(\zeta) = \int_{\zeta} {\widehat {\cal O}}^{(1)}_A,  \qquad   W_A(\Sigma) = \int_{\Sigma} {\widehat {\cal O}}^{(2)}_A,
\label{wa}
\ee
where
\be
\{Q, W_A(P) \} = \{Q, W_A(\zeta) \} = \{Q, W_A(\Sigma) \} = 0.
\ee
In addition, we also have the descent relations
\be
d{\widehat {\cal O}}^{(0)}_A  = \{Q, {\widehat {\cal O}}^{(1)}_A \}, \qquad  d{\widehat {\cal O}}^{(1)}_A  = \{Q, {\widehat {\cal O}}^{(2)}_A \}, \qquad  d{\widehat {\cal O}}^{(2)}_A=0.
\label{descent}
\ee
In the above relations,
\be
{\widehat {\cal O}}^{(1)}_A  \in \Gamma (\Omega^1_{\Sigma} \otimes (\Omega^{ch}_X)^{\mathfrak t \geq} \otimes \langle \phi^a \rangle),   \qquad  {\widehat {\cal O}}^{(2)}_A  \in  \Gamma (\Omega^2_{\Sigma} \otimes (\Omega^{ch}_X)^{\mathfrak t \geq} \otimes \langle \phi^a \rangle), \
\ee
and so from (\ref{wa}), we find that $W_{A}(P), W_{A}(\zeta) \ \textrm{and} \ W_{A}(\Sigma)$ will be given by global sections of $(\Omega^{ch}_X)^{\mathfrak t \geq} \otimes \langle \phi^a \rangle$. Moreover, since $W_{A}(P), W_{A}(\zeta) \ \textrm{and} \ W_{A}(\Sigma)$ are $Q$-closed, they will correspond to classes in  the chiral equivariant cohomology ${\bf H}_{T^d} (\Omega^{ch}_X)$. From the descent relations in (\ref{descent}), we also find that with respect to the $Q$-cohomology and therefore ${\bf H}_{T^d} (\Omega^{ch}_X)$, the operators ${\widehat {\cal O}}^{(0)}_A$, ${\widehat {\cal O}}^{(1)}_A$ and ${\widehat {\cal O}}^{(2)}_A$ can be viewed as $d$-closed forms on $\Sigma$ (since their exterior derivatives on $\Sigma$ are $Q$-exact and therefore trivial in $Q$-cohomology).

\vspace{0.4cm}{\noindent \it {Relation to the Classical Equivariant Cohomology of $X$}}

Consider the operator $W_{A_l} (\gamma_l)$, where $A_l$ is associated with the operator ${\cal O}_{A_l}$ in (\ref{OA}) that is of degree $n_l$ in the fields $\psi^i$, and $\gamma_l$ is a homology cycle on $\Sigma$ of dimension $t_l$. Notice that $W_{A_l} (\gamma_l)$ generalises the operators $W_A(P)$, $W_A(\zeta)$ and $W_A(\Sigma)$ above.  Now consider a general correlation function of $s$ such operators:
\be
{Z((A_1, \gamma_1, \dots, (A_s, \gamma_s))} =  {\left \langle \ {\Pi_{l=1}^s W_{A_l}(\gamma_l)} \ \right \rangle}_0.
\ee
This can be explicitly written as
\be
{Z((A_1, \gamma_1, \dots, (A_s, \gamma_s))} = \int {\cal D}X \ e^{-S_{\textrm{gauged}}}\cdot {\Pi_{l=1}^s W_{A_l}(\gamma_l)},
\label{cf}
\ee
where ${\cal D}X$ is an abbreviated notation of the path integral measure ${\cal D} A \cdot {\cal D}\phi \cdot {\cal D} \psi \cdot {\cal D} \phi^a \cdot {\cal D} \psi^a$ over all inequivalent field configurations.

As a relevant digression at this point, let us present an argument made in sect. 5 of \cite{mirror manifolds}. Consider an arbitrary quantum field theory, with some function space $\cal E$ over which one wishes to integrate. Let $F$ be a group of symmetries of the theory. Suppose $F$ acts freely on $\cal E$. Then, one has a fibration ${\cal E} \to {\cal E}/F$, and by integrating first over the fibres of this fibration, one can reduce the integral over $\cal E$ to an integral over ${\cal E}/F$. Provided one considers only $F$-invariant observables $\cal O$, the integration over the fibres will just give a factor of $\textrm{vol}(F)$ (the volume of the group $F$):
\be
\int_{\cal E} e^{-S} {\cal O} = \textrm{vol} (F) \cdot \int_{{\cal E}/F} e^{-S} {\cal O}.
\label{fpt}
\ee
Since $G$ is a freely-acting gauge symmetry of our sigma model, and since the $W_{A_l} (\gamma_l)$'s are $G$-invariant operators, we can apply the above argument to our case where $F=G$, and ${\cal O} = {\Pi_{l=1}^s W_{A_l}(\gamma_l)}$. Thus, for the correlation function path integral in (\ref{cf}), the integration is done over fields modulo gauge transformations, that is, over orbits of the gauge group. This observation will be essential below.

Applying the same argument with $F$ being the group of supersymmetries generated by $Q$, and $\cal O$ being the product of $Q$-invariant operators ${\Pi_{l=1}^s W_{A_l}(\gamma_l)}$, we learn that the path integral in (\ref{cf}) will localise onto $Q$-fixed points only \cite{mirror manifolds}, that is, from (\ref{gauge tx first})-(\ref{gauge tx last}), onto the field configurations whereby $\psi^a_z = \psi^a_{\bar z} = 0$, $\phi^a = 0$, $\partial_z \phi^a = \partial_{\bar z} \phi^a = 0$, and $\partial_{\bar z} \phi^i = \partial_z \phi^{\bar i} =0$. Hence, the path integral localises onto the moduli space of holomorphic maps $\Phi$ modulo gauge transformations. As explained earlier, one considers only degree-zero maps in perturbation theory. Since the space of holomorphic maps of degree-zero is the target space $X$ itself, we find that for the path integral in (\ref{cf}), one simply needs to integrate over the quotient space $X/G$.

As pointed out earlier,    the $W_{A_l}(\gamma_l)$'s represent weight-zero classes in the chiral equivairant cohomology ${\bf H}_{T^d} (\Omega^{ch}_X)$. Granted that as claimed in \cite{andy1,andy2}, one has a mathematically consistent isomorphism between the weight-zero classes of ${\bf H}_{T^d} (\Omega^{ch}_X)$ and the classical equivariant cohomology $H_G(X)$, it will mean that there is a one-to-one correspondence between the $W_{A_l}(\gamma_l)$'s and the elements of $H_G(X)$. Since the $G$-action on $X$ is freely-acting, that is, the quotient space $X/G$ is a smooth manifold, we will have $H_G(X) = H(X/G)$, where $H(X/G)$ is just the de Rham cohomology of $X/G$. This means that the correlation function in (\ref{cf}) will be given by
\be
{Z((A_1, \gamma_1, \dots, (A_s, \gamma_s))} = \int_{X/G} {\cal W}_{A_1} \wedge {\cal W}_{A_2} \wedge \dots {\cal W}_{A_s},
\label{cftopo}
\ee
where ${\cal W}_{A_i}$ is just an appropriate, globally-defined differential form in the de Rham cohomology of $X/G$ corresponding to the physical operator $W_{A_i} (\gamma_i)$, such that ${\sum_{i=1}^s \textrm{degree} ( {\cal W}_{A_i}}) = \textrm{dim} (X/G)$. Notice that the right-hand side of (\ref{cftopo}) is an intersection form and is thus a topological invariant of $X/G$ and hence $X$, for a specified gauge group $G$ that is freely-acting. This is consistent with the earlier physical observation that ${Z((A_1, \gamma_1, \dots, (A_s, \gamma_s))}$ is a topological invariant of $X$.  Therefore, we conclude that the mathematical isomorphism between the weight-zero classes of ${\bf H}_{T^d} (\Omega^{ch}_X)$ and the classical equivariant cohomology $H_G(X)$, is likewise consistent from a physical viewpoint via the interpretation of the chiral equivairant cohomology as the spectrum of ground operators in the half-twisted gauged sigma model.

\newsubsection{A Topological Chiral Ring and the de Rham Cohomology Ring of $X/G$}

Recall from sect. 3.5 that the local operators of the perturbative half-twisted gauged sigma model will span a holomorphic chiral algebra. In particular, one can bring two local operators close together, and their resulting OPE's will have holomorphic structure coefficients. The ${\widehat {\cal O}}^{(0)}_{A_i}$'s, or rather $W_{A_i}(P)$'s, are an example of such local, holomorphic operators. By holomorphy, and the conservation of scaling dimensions and $(g_L, g_R)$ ghost number, the OPE of  these  operators take the form
\be
{W_{A_i}(z) W_{A_j}(z')} = {\sum_{g_k =  g_i + g_j}  { {C^k_{ij} \ W_{A_k}(z')} \over {(z- z')^{h_i + h_j -h_k} } } },
\label{OPEij}
\ee
where $z$ and $z'$ correspond to the points $P$ and $P'$ on $\Sigma$, and the $h_{\alpha}$'s are the holomorphic scaling dimensions of the operators. We have also represented the $(g_L, g_R)$ ghost numbers of the operators $W_{A_i}(z)$, $W_{A_j}(z)$ and $W_{A_k}(z)$ by $g_i$, $g_j$ and $g_k$ for brevity of notation. Here, $C^k_{ij}$ is a structure coefficient that is (anti)symmetric in the indices. Since $W_{A_i}(z)$ and $W_{A_j}(z)$ are ground operators of dimension $(0,0)$, i.e., $h_i = h_j =0$, the OPE will then be given by
\be
{W_{A_i}(z) W_{A_j}(z')} = {\sum_{g_k =  g_i + g_k} { C^k_{ij} \ {W_{A_k}(z')} \over {(z- z')^{-h_k}}}}.
\label{OPEk}
\ee
Notice that the RHS of (\ref{OPEk}) is only singular if $h_k < 0$. Also recall that all physical operators in the $Q_R$-cohomology cannot have negative scaling dimension, that is, $h_k \geq 0$. Hence, the RHS of (\ref{OPEk}), given by $(z-z')^{h_k} W_{A_k}(z')$, is non-singular as $z \to z'$, since a pole does not exist. Note that $(z-z')^{h_k}  W_{A_k}(z')$ 
must also be annihilated by $Q_R$ and be in its cohomology, since this is true of $W_{A_i}(z)$ and $W_{A_j}(z')$ too. In other words, we can write $W_{A_k} (z, z') = (z-z')^{h_k} W_{A_k}(z')$, where $W_{A_k} (z, z')$ is a dimension $(0,0)$ operator that represents a $ Q_R$-cohomology class.  Thus, we can express the OPE of the ground operators as
\be
{W_{A_i}(z) W_{A_j}(z')} = {\sum_{g_k =  g_i + g_j}  C^k_{ij} \ W_{A_k}(z , z')}.
\label{OPEgndgauge}
\ee
Since the only holomorphic functions without a pole on a Riemann surface are constants, it will mean that the operators $W_{A_k}(P)$, as expressed in the OPE above, can be taken to be independent of the coordinate `$z$' on $\Sigma$. Hence, they are completely independent of their insertion points and the metric on $\Sigma$. Therefore, we conclude that the ground operators of the chiral algebra $\cal A$ of the sigma model define a $\it{topological}$ chiral ring via the OPE
\be
{ W_{A_i} W_{A_j}} = {\sum_{g_k =  g_i + g_j}  C^k_{ij} \ W_{A_k}}.
\label{OPEgndgauge1}
\ee

Now, consider the following two-point correlation function
\be
\eta_{ij} = {\langle W_{A_i} W_{A_j} \rangle}_0.
\label{2point}
\ee
Next, consider the three-point correlation function
\be
{\langle W_{A_i} W_{A_j} W_{A_k} \rangle}_0 = {\langle W_{A_i}( W_{A_l} C^l_{jk}) \rangle}_0 = {\langle W_{A_i}W_{A_l}\rangle}_0 \ C^l_{jk},
\label{3point}
\ee
where we have used the OPE in (\ref{OPEgndgauge1}) to arrive at the first equality above. Thus, if we let
\be
{\langle W_{A_i} W_{A_j} W_{A_k} \rangle}_0 = C_{ijk},
\ee
from (\ref{2point}) and (\ref{3point}), we will have
\be
C_{ijk} = \eta_{il} C^l_{jk}.
\label{deter}
\ee
From the discussion in the previous subsection, we find that
\be
C_{ijk} = \int_{X/G} {\cal W}_{A_i} \wedge {\cal W}_{A_j} \wedge {\cal W}_{A_k}
\ee
and
\be
\eta_{il} = \int_{X/G} {\cal W}_{A_i}\wedge {\cal W}_{A_l},
\ee
that is, $\eta_{il}$ and $C_{ijk}$ correspond to the intersection pairing and structure constant of the de Rham cohomology of $X/G$ respectively. Therefore,  one can see that the two-point correlation function of local ground operators at genus-zero defined in (\ref{2point}), and the structure coefficient $C^l_{jk}$ of the topological chiral ring in (\ref{OPEgndgauge1}), will, together with (\ref{deter}), determine the de Rham cohomology ring of $X/G$ completely.

\newsubsection{Results at Arbitrary Values of the Sigma Model Coupling}

From (\ref{Sgauged}) and (\ref{Vgauged}), we see that the Lagrangian in (\ref{lgauged})  of the half-twisted gauged sigma model, can be written as
\be
{\cal L}_{\textrm{gauged}} = \{Q_L, V_{\textrm{gauged}} \} + \{Q_R, V_{\textrm{gauged}} \},
\ee
where $V_{\textrm{gauged}}$ is given explicitly by
\be
V_{\textrm{gauged}} = i g_{i \bar j} ( \psi^i_{\bar z} D_z \phi^{\bar j} + \psi^{\bar j}_z D_{\bar z} \phi^i - {1\over 2} \psi^{\bar j}_z H^i_{\bar z} - {1\over 2} \psi^i_{\bar z}H^{\bar j}_z).
\ee
Consequently, one can see that any change in the metric $g_{i \bar j}$ will manifest itself as a $Q_R$-exact and a $Q_L$-exact term. The $Q_R$-exact term is trivial in $Q_R$-cohomology, while the $Q_L$-exact term is trivial in $Q_L$-cohomology. Therefore, arbitrary changes in the metric can be ignored when analysing the subset of operators of the half-twisted gauged sigma model that are also in the $Q_L$-cohomology. In particular, one can move away from the infinite-volume limit to a large but finite-volume regime of the sigma model (where worldsheet instanton effects are still negligible), and the above discussion on the operators of the $Q$-cohomology will not be affected. Thus, the interpretation of the chiral equivariant cohomology as the ground operators of the half-twisted gauged sigma model hold at arbitrarily small values of the coupling constant and hence, to all orders in perturbation theory. Likewise, this will also be true of the physical verification of the isomorphism between the weight-zero subspace of the chiral equivariant cohomology and the classical equivariant cohomology of $X$, and the relation of the intersection pairing and structure constant of the de Rham cohomology ring of $X/G$ to the two-point correlation function and structure coefficient of the topological chiral ring, whereby their validity rests upon arguments involving operators in the $Q$-cohomology.

\newsection{Concluding Remarks}

In this paper, we have furnished a purely physical interpretation of the chiral equivariant cohomology defined by Lian and Linshaw \cite{andy1} in terms of a two-dimensional sigma model. In particular, for a locally-free and abelian $G$-action such as $G=T^d$, the chiral equivariant cohomology of a $G$-manifold $X$ will correspond to the sub-spectrum of ground operators of the half-twisted $G$-gauged sigma model which are also in the $Q_L$- and $(Q_L +Q_R)$-cohomology. Via this sigma model interpretation, the vanishing of positive weight classes in the chiral equivariant cohomology can be attributed to the simple  physical observation that the holomorphic and anti-holomorphic stress tensors of the model are $Q_L$- and $Q_R$-closed respectively; hence, any admissible operator that is both in the $Q_L$- and $Q_R$-cohomology at the same instant must be of weight $(0,0)$. Moreover, we have also verified, from a purely physical perspective using the topological invariance of the  correlation function of local and non-local operators, the validity of identifying the weight-zero subspace of the chiral equivariant cohomology with the classical equivariant cohomology of $X$. Last but not least, we have also demonstrated that the de Rham cohomology ring of $X/G$ can be determined fully from the two-point correlation function of local ground operators which span the chiral algebra, and the structure coefficient of the topological chiral ring generated by these local operators. Hopefully, the math-physics connection elucidated in this present work can bring about further progress in either fields through an application of the physical and mathematical insights that it may have offered.

What remains to be explored is the case when the abelian $G$-action has fixed-points, that is, when the target space of the half-twisted gauged sigma model is a singular orbifold. According to the results of \cite{andy2}, there will be non-vanishing classes of positive weights in the corresponding chiral equivariant cohomology. Again, it would be interesting and probably useful to understand this from a purely physical perspective.

Finally, it would also be interesting to provide a physical interpretation of the chiral equivariant cohomology of $X$ when $G$ is a non-abelian group. From the mathematical construction in \cite{andy1}, we find that the chiral Cartan complex in the Cartan model of the chiral equivariant cohomology, is now a tensor product of the horizontal subalgebra of the semi-infinite Weil algebra and the chiral de Rham complex. This is in contrast to the small chiral Cartan complex discussed in this paper, which is just a tensor product of $\langle \phi^a \rangle$ and the chiral de Rham complex. The work of Getzler \cite{getzler}, which aims to examine the analogy between equivariant cohomology and the  topological string, involves the semi-infinite Weil algebra. This seems to suggest that perhaps one should consider a topological string extension of the half-twisted gauged sigma model, that is, to consider coupling the present model to two-dimensional worldsheet gravity in a BRST-invariant fashion, such that one will need to integrate over the space of all inequivalent worldsheet Riemann surfaces in any path integral computation. The resulting model may just provide a physical interpretation of the chiral equivariant cohomology in the non-abelian case. We hope to explore this consideration elsewhere in a future publication.

\vspace{0.5cm}
\hspace{-1.0cm}{\large \bf Acknowledgements:}\\
I would like to take this opportunity to thank B.E. Baaquie, B.H. Lian, A. Linshaw and E. Witten for providing their expert opinion on various issues in the course of our discussions.
\newline


\vspace{-1.0cm}




\begin{thebibliography}{99}


\bibitem{andy1}

Bong H. Lian, Andrew R. Linshaw, ``Chiral Equivariant Cohomology I'', Adv. Math. {\bf{209}}, 99-161 (2007), [arXiv: math.DG/0501084].

\bibitem{andy2}

Bong H. Lian, Andrew R. Linshaw, Bailin Song, ``Chiral Equivariant Cohomology II'', Trans. Amer. Math. Soc. 360 (2008), 4739-4776. [arXiv: math.DG/0607223].




\bibitem{MSV1}


F. Malikov, V. Schechtman, and A. Vaintrob, ``Chiral De
Rham Complex'', Comm. Math. Phys. {\bf{204}} (1999) 439-473, [arXiv:math.AG/9803041].

\bibitem{MSV2}

F. Malikov and V. Schechtman, ``Chiral de Rham Complex II'', preprint, [arXiv:math.AG/9901065].




\bibitem{Bo}

L. Borisov, ``Vertex Algebras and Mirror Symmetry", Comm. Math. Phys. {\bf{215}} (2001) 517-557.


\bibitem{BL}

L. Borisov, A. Libgober, ``Elliptic Genera and Applications to Mirror Symmetry'', Inventiones Math. {\bf{140}} (2) (2000), 453-485.

\bibitem{BL1}



L. Borisov, A. Libgober, ``Elliptic Genera of Singular Varieties'', Duke Math. J. {\bf{116}}, 2 (2003), 319-351, [arXiv:math.AG/0007108].


\bibitem{BL2}

L. Borisov, A. Libgober, ``Elliptic Genera of Singular Varieties, Orbifold Elliptic Genus and Chiral
De Rham Complex'',  Mirror Symmetry IV (Montreal, QC, 2000), 325--342, AMS/IP Stud. Adv. Math., 33, Amer. Math. Soc., Providence, RI, 2002. [arXiv:math.AG/0007126].


\bibitem{Ka}

A. Kapustin, ``Chiral de Rham Complex and the Half-Twisted Sigma-Model'', preprint, [arXiv:hep-th/0504074].

\bibitem{MC}

M.-C. Tan, "Two-Dimensional Twisted Sigma Models and the Theory of Chiral Differential
Operators", Adv. Theor. Math. Phys. {\bf{10}}, 759 (2006), [arXiv: hep-th/0604179].

\bibitem{n=2}

E. Witten, ``Phases of N = 2 Theories in Two Dimensions", Nucl. Phys. {\bf{B403}} (1993)
159-222, [arXiv:hep-th/9301042].

\bibitem{mirror manifolds}

E. Witten, ``Mirror Manifolds And Topological Field Theory", in Essays On Mirror
Manifolds, ed. S.-T. Yau (International Press, Hong Kong, 1992), [arXiv:hep-th/9112056].

\bibitem{bagger}

J. Wess and J. Bagger, ``Supersymmetry And Supergravity'', Princeton University Press (second edition, 1992).


\bibitem{CDO}

Ed Witten, ``Two-Dimensional Models with (0,2) Supersymmetry: Perturbative Aspects", Adv. Theor. Math. Phys. {\bf{11}}, 1 (2007), [arXiv:hep-th/0504078].




\bibitem{GSW2}

M. Green, J.H. Schwarz and E. Witten, ``Superstring Theory, Vol II." (Cambridge, Cambridge
University Press, 1987).





\bibitem{sternberg}

V. Guillemin and S. Sternberg, ``Supersymmetry and Equivariant de Rham Theory'',
Springer, 1999.


\bibitem{Bott}

Bott, R., Tu, L.: ``Differential Forms in Algebraic Topology". Berlin, Heidelberg, New York: Springer 1982.


\bibitem{getzler}

E. Getzler, ``Two-dimensional topological gravity and equivariant cohomology'', Commun.
Math. Phys. {\bf{163}} (1994) 473-489, [arXiv:hep-th/9305013].


\end{thebibliography}
\end{document}